\documentclass[11pt,a4paper]{article}
\usepackage{amssymb,amsmath}
\usepackage{graphicx}
\usepackage{fancyhdr}
\usepackage{xcolor}
\usepackage{lineno}
\usepackage{url}
\makeatother
\renewcommand{\headrulewidth}{0pt}
\usepackage{overcite}

\begin{document}

\newcommand{\mib}[1]{\mbox{\boldmath$#1$}}
\newcommand{\tx}[1]{\textrm{\scriptsize #1}}

\newcommand{\rev}[1]{\textcolor{black}{#1}}    
\newcommand{\editor}[1]{\textcolor{black}{#1}}  


\begin{center}
\editor{{\Large Jupiter's cloud-level variability triggered by torsional oscillations in the interior}} \\
\vspace{10mm}
Kumiko Hori$^{1*}$, Chris A. Jones$^2$, Arrate Antu\~{n}ano$^3$, Leigh N. Fletcher$^4$, and Steven M. Tobias$^2$  \\
\vspace{10mm}
$^1$Graduate School of System Informatics, Kobe University, Kobe, Japan.  \\
$^2$Department of Applied Mathematics, University of Leeds, Leeds, UK.\\
$^3$Departamento F\'{i}sica Aplicada, Escuela de Ingenier\'{i}a de Bilbao, Universidad del Pa\'{i}s Vasco UPV/EHU, Spain.\\ 
$^4$School of Physics and Astronomy, University of Leicester, Leicester, UK.\\
$^*$The corresponding author. \\
\end{center}

\renewcommand{\abstractname}{}

\begin{abstract}
\noindent\rule[.6ex]{\linewidth}{.06ex}

\vspace{-5mm}
\editor{
\subsection*{Abstract}
}
\vspace{5mm}

Jupiter's weather layer exhibits long-term and quasi-periodic cycles of meteorological activity \rev{that can completely change the appearance of its belts and zones}. 
There are cycles with intervals from 4 to 9 years, dependent on the latitude, which were 
detected in 5\,$\mu$m radiation,
 which provides a window \rev{into the cloud-forming regions of the troposphere}; 
 however, the origin of these cycles has been \editor{a mystery.
Here} 
\rev{we propose that 
 magnetic torsional 
 oscillations/waves arising from the dynamo region
 could modulate the heat transport and hence be ultimately responsible for the variability of the 
 tropospheric banding. 
These axisymmetric waves are magnetohydrodynamic waves influenced by the rapid rotation,
 which have been detected in Earth's core, and have been recently suggested to exist in Jupiter by 
 the observation of magnetic secular variations by Juno}. 
Using the magnetic field model \rev{JRM33}, together \editor{a} the density distribution model, 
we compute the expected speed of these waves. 
For the \editor{waves} excited by variations in the zonal jet flows, their wavelength can be estimated from the width of the alternating jets, yielding waves with a half period of \rev{3.2-4.7} years in 14-23$^\circ$N,  
 consistent with the intervals \rev{with the cycles of variability of Jupiter's North Equatorial Belt and North Temperate Belt} identified in the \rev{visible and} infrared observations. 
\editor{The nature of these waves, including the wave speed and the wavelength, is revealed by a data-driven technique, dynamic mode decomposition, applied to the spatio-temporal data for 5\,$\mu$m emission. 
Our results imply}
 that exploration of these magnetohydrodynamic waves may provide a new window to \rev{the origins of quasi-periodic patterns in Jupiter's tropospheric clouds and to} the internal dynamics and the dynamo of Jupiter.

\noindent\rule[.6ex]{\linewidth}{.06ex}
\end{abstract}

\editor{
\section*{Main text}
}
\vspace{10mm}


Significant variability has been seen in Jupiter's cloud deck for over 100 years\cite{R95,F17} with colour changes, brightening events, and upheavals 
\editor{owing} to outbreaks of convective plumes\cite{S86,Getal00}~\hspace{-2mm}.
These upheavals occur \rev{
  on the jet streams
  and have significant consequences for the temperature, composition, and cloud distributions observed in Jupiter's belts and zones, where belts (zones) refer to latitudinal bands that are cyclonic (anticyclonic) and dark (bright) \editor{at visible wavelengths}.} 
This variability can also be detected by infrared observations at 5\,$\mu$m\cite{GLS69,W69}~\hspace{-1.5mm}.
At this wavelength, Jupiter's infrared emission from the \rev{1-7 bar} region is modulated by overlying aerosol opacity, both from condensed clouds and aerosols\rev{\cite{GFI15,BWPA15}}~\hspace{-2mm}.
Belts are typically bright, being free from clouds, whereas zones are typically dark and cloudy.

Records of infrared observations at the 5\,$\mu$m atmospheric window now cover \rev{more than three Jovian years (35.6 Earth years)} at the low-mid latitudes\cite{F17}~\hspace{-1.5mm}.
These observations sense the cloud-free atmosphere down to the \rev{7 bar} level, but the changes in the radiance are likely associated with opacity variations in the $\mathrm{NH_3}$ ice clouds and/or the solid $\mathrm{NH_4SH}$ forming region 
 at the 1-2 bar level\cite{WBetal04}~\hspace{-1.5mm}. 
\editor{The zonally averaged brightness
 shows} cyclic changes with periods in the range 4-9 years\cite{AFetal18,AFetal19}~\hspace{-1.5mm}.
The periods identified, and their corresponding latitude bands, are shown in the first two columns of Table~\ref{table:periods}. 
These periods are similar to that of 
\editor{the quasi-quadrennial oscillation, seen in 
the equatorial stratospheric temperature. This 
variation} is believed to have a similar origin to that of the quasi-biennial oscillation in the Earth's atmosphere, which is generally thought to be forced by upward propagating gravity waves driven by convection. 
However, the periods seen in the 5\,$\mu$m wavelength originate from much deeper than the Jovian stratosphere, and are also seen in the mid-latitudes, not just in the equatorial region\editor{.}
The origins of the periodic variations are not well known, and have been thought to be the result of either top-down control (e.g., related to oscillations in the stratosphere) or bottom-up control as a result of convective processes in the troposphere\rev{\cite{F17,AFetal19}~\hspace{-1.5mm}.
No theoretical models have accounted for the observed period 
between upheavals.}

\vspace{5mm}

\subsection*{Torsional oscillation speed in the interior} 

\begin{figure}[ht]
  \includegraphics{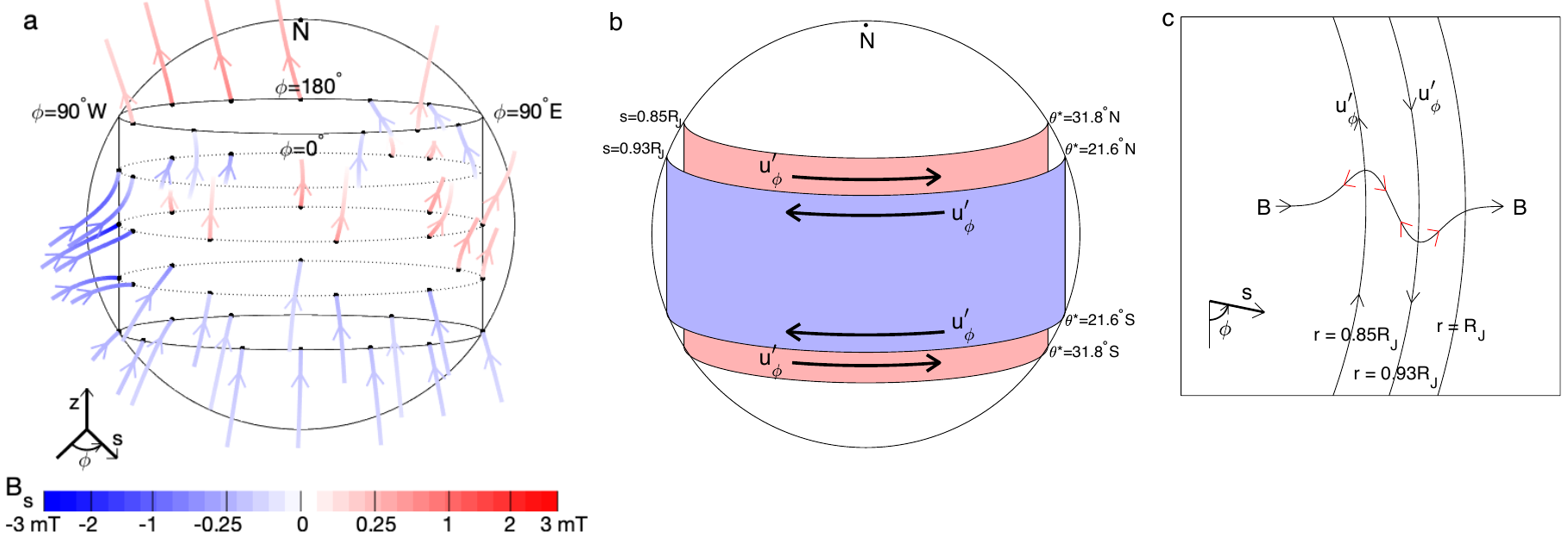}
\caption{\editor{Torsional oscillation in Jupiter.} (a) Jupiter's equilibrium magnetic field lines outside the tangent cylinder $s=0.85R_\tx{J}$.
  Blue field lines enter this tangent cylinder and red field lines leave the tangent cylinder at the black dots.
  The tangent cylinder has height $\pm h=R_\tx{J}(1-0.85^2)^{1/2}$, and field lines passing through $z=-h$, $-h/2$, $0$, $h/2$ and $h$ at $s=0.85\,R_\tx{J}$ are shown.
  The colour scale shows $B_s$, the $s$-component of the magnetic field \mib{B} in mT.
  Note the strong negative $B_s$ near the \lq great blue spot' at longitude $90^\circ$\,W close to the equator.
  This spot greatly enhances the torsional wave speed. Magnetic field data is based on the model JRM33\cite{Cetal21}~\hspace{-1.5mm}.
  (b) Schematic diagram of a torsional oscillation. The fluid speed is constant on coaxial cylinders, so $u_\phi^\prime$ is a function of $s$ and $t$ only.
  The cylinders with $s=0.85R_\tx{J}$ and $s=0.93R_\tx{J}$ are shown, corresponding to latitudes $\theta^* =31.8^\circ$ and $21.6^\circ$.
  (c) Schematic diagram of the effect of a torsional oscillation on the equilibrium magnetic field in the equatorial plane.
  The equilibrium field shown is outward in the radial direction, but it is distorted by the cylindrical flow, which tries to move the field along with the flowing fluid, as shown.
  The magnetic field provides a tension, as would a stretched string, and this provides a restoring force, as indicated by the red arrows, that allows the cylinders to oscillate.
  If the field is not aligned with the $s$-direction, only the $s$-component of $\mib{B}$ provides the restoring force.}
 \label{fig:torsional}
\end{figure}

\rev{We here propose that torsional oscillations give rise to a shear that disrupts the slow convective flows in the deep interior that carry the heat flux towards the visible troposphere.
The torsional waves themselves can only exist in the electrically conducting region of Jupiter, which may effectively span up to $\sim$96\% of the planet's radius \editor{$R_\tx{J}$}(\editor{details below}), so they are 
\editor{excluded from the equatorial region,
$\sim$16$^\circ$N/S,}
all latitudes being planetocentric.  
However, if they disrupt the radial heat flux at depth, they could be responsible for changes in the emerging heat flux even at low latitudes.} 
This raises the possibility that the periods observed in the 5$\mu$m signal\cite{AFetal19} are the result of disturbances triggered by torsional oscillations of the magnetic field deep in Jupiter's interior\cite{Bra70,HTJ19}~\hspace{-1.5mm}.
In a rapidly rotating body like Jupiter, fluid disturbances that bend the lines of planetary vorticity parallel to the rotation axis have periods of days rather than years,
but torsional oscillations correspond to rotational fluid motions in the longitudinal $\phi$ direction in which the velocity is constant on coaxial cylinders.
These motions can have periods of years rather than days, the period being determined by the magnetic coupling between neighbouring cylinders.
Torsional oscillations are illustrated in Figure~\ref{fig:torsional}.
We use two coordinate systems, spherical polars $(r, \theta, \phi)$ \editor{with 
$\theta$} being the co-latitude, and cylindrical polars $(s, \phi, z)$ \rev{where the $z$ coordinate is parallel to the rotation axis.
For} simplicity \rev{we consider a spherical planet}, \rev{ignoring the small errors that will occur because of Jupiter's oblateness.}
A cylinder of radius $s$ cuts the surface at latitude $\rev{\theta^*}$
 given by $\cos(\rev{\theta^*}) = \sin(\theta) = s/R_\tx{J}$, which provides a simple correspondence between each latitude (north or south) and a particular $s$-value.
\editor{Torsional waves propagate in both \rev{the positive and negative $s$-directions} 
with Alfv\'en speed} $U_\tx{A} = (\langle \overline{B_s^2} \rangle/\rev{\mu_0} \langle \overline{\rho} \rangle )^{1/2}$ at each $s$ locally
\editor{(}where $B_s$, $\rho$, and $\rev{\mu_0}$ are the $s$-component of the magnetic field, the density, and the magnetic permeability; $\langle \overline{f} \rangle$ denotes the $z$- and $\phi$- mean of any arbitrary function $f$\editor{)},
so that both travelling and standing waves are possible, 
 depending on how the waves are forced \editor{(Methods)}.

Figure~\ref{fig:torsional}a shows the magnetic field lines outside the cylinder with $s=0.85 R_\tx{J}$. 
\rev{The radius $r=0.85R_\tx{J}$ may be regarded as an interface to the dynamo,} and the colour of each line gives the local value of $B_s$. 
The field is constructed from 
\editor{the Gauss coefficients tabulated in the Juno data
JRM33 magnetic field model\rev{\cite{Cetal21}}~\hspace{-1.5mm}.}
The field is assumed to be \rev{potential} for Fig.~\ref{fig:torsional}a, i.e. electrical currents there are negligible (\editor{Methods}).
\rev{Using a realistic density profile\cite{FBLNBWR12}~\hspace{-1.5mm}, the wave speed $U_\tx{A}$ as a function of $s$ is calculated. \rev{The profile is demonstrated in Supplementary Figure~1, and}  
 the selected values of $s$ and the corresponding $U_\tx{A}$ are listed} in columns 3 and 4 of Table~\ref{table:periods}.
\rev{\editor{Whilst the recent magnetic model\cite{Cetal21} suggested a dynamo top located at $r= 0.81 R_\tx{J}$,
simulations\cite{TJ20}} found that dynamo generated currents start to affect \rev{the} magnetic field significantly inside $r=0.9R_\tx{J}$,
so that outside $r=0.9R_\tx{J}$ the torsional wave speed can be computed fairly reliably from a \rev{current-free} approach.}
Inside $r=0.9R_\tx{J}$, the waves still exist but their wave speed becomes progressively more uncertain as the depth increases.
The potential field rises rapidly below $r=0.9R_\tx{J}$, but simulations suggest that the dynamo generated field rises more slowly, so the wave speed computed from the potential field at $0.85R_\tx{J}$ is likely to be somewhat faster than the true wave speed.
Figure~\ref{fig:torsional}b illustrates the oscillation of zonal velocity, 
\editor{$\langle \overline{u_\phi^\prime} \rangle$}, that gives rise to the torsional waves,
and Figure~\ref{fig:torsional}c shows how the Lorentz force from the perturbed magnetic field provides the restoring force that creates the waves.

A notable feature of Jupiter's magnetic field is the \lq great blue spot' region located near the equator around longitude 90$^\circ$\,W\cite{Metal18}~\hspace{-1.5mm}.
\editor{The field here} is strong and nearly radially inward, so there is a particularly high value of $B_s^2$\editor{, making} a major contribution to $\langle \overline{B_s^2} \rangle$.
\editor{The} colour scale on Fig.~\ref{fig:torsional}a is nonlinear so that the other field lines are not too faint. 
If the magnetic field were exactly dipolar, the field in the region outside $s=0.85R_\tx{J}$ would predominantly be in the $z$ direction, and \editor{the wave speed} would be much less than it actually is.

In the outer regions of the planet the electrical conductivity starts to fall dramatically as the pressure is insufficient to create metallic hydrogen.
Torsional waves, which rely on induced currents, therefore become strongly damped. 
Using the H/He model conductivities\cite{FBLNBWR12}~\hspace{-1.5mm}, we estimate the depth at which critical damping occurs to be $0.94R_\tx{J}$, though it is possible that it could be somewhat further out,
 say $0.96 R_\tx{J}$\editor{,} 
where the coaxial cylinders intersect the surface at latitudes $\pm 16^\circ$, if heavier elements enhance the electrical conductivity.
\rev{Torsional waves further out than this would be heavily damped, so when calculating \editor{the speed} $(\langle \overline{B_s^2} \rangle/\mu_0 \langle \overline{\rho} \rangle )^{1/2}$, we only integrate the field out to $r= 0.96 R_\tx{J} \equiv r_\text{diff}$ (\editor{Methods}). 
If the mechanism by which torsional waves affect the 5\,$\mu$m radiation \editor{was} local, i.e. it \editor{was} only cylindrical fluctuations extending to the surface,
we would not expect to see any activity outside the $\pm 16^\circ$ region. 
However, deep torsional waves may modulate the heat flux coming out of Jupiter at a depth of \editor{$\sim$3500\,km} below the surface, $0.95 R_\text{J}$, where there is sufficient electrical conductivity, and then it could be possible for this modulated flux to reach the surface even in the equatorial regions. 
The global nature of the 5\,$\mu$m variability\cite{AFetal19} suggests 
that a deep origin is plausible.}

\vspace{5mm}

\subsection*{Torsional oscillation period at varying latitudes} 

To convert the wave speeds into temporal periods we need to know the wavelength in the latitudinal direction, which can then be converted into a wavelength in the $s$-direction
and a local wavenumber $k$. Then we can calculate the frequency, $\omega_\tx{TW}= U_\tx{A} k$, and the period, $T = 2\pi/\omega_\tx{TW}$, at the given latitudes.
The spatial scale is determined by the excitation mechanism.
We suspect torsional waves \rev{in a gas giant} will predominantly be forced by fluctuations in the Reynolds stress, primarily due to the prominent zonal flows \cite{HTJ19}(\editor{Methods and Supplementary Figures~2a-e}),
 in contrast to torsional waves in the Earth's core, which are most likely driven by magnetic stresses\cite{TJT15}~\hspace{-1.5mm}. 
\rev{While the latitudes of the extrema of the zonal flows, of the order of 100\,m/sec, are fairly constant in time,
 their magnitude varies by around 10\,m/sec\cite{TOLL17,Wetal20}}~\hspace{-1.5mm}.
It is now believed that the time-averaged zonal winds extend
\rev{down to a few thousands of kilometres depth, reaching only a small value at that depth\cite{KGHetal18,GK21}~\hspace{-1.5mm}.}
\rev{This steady component of the zonal flow varies in the $z$-direction, likely in balance with temperature fluctuations (thermal wind balance\cite{KGHetal18}), a balance set up over a very long time-scale. 
Decadal timescale torsional oscillations are too fast to set up a thermal wind, so we expect them to be geostrophic ($z$-independent) outside of stably stratified regions, as shown in Fig.~\ref{fig:torsional}.}
\rev{At still greater depth, Juno-detected magnetic secular variation suggests a velocity of 4\,cm/sec at a depth of 3500 km, $\sim$0.95$R_\tx{J}$\cite{Metal19,Cetal21}~\hspace{-1.5mm}. 
However, some of this 4\,cm/sec might be accounted for by errors in the system III rotation rate, so that the true velocity could be as little as 1\,cm/sec\cite{Betal22}~\hspace{-1.5mm}.  
There is also some evidence that this flow is not steady, but is decreasing on a decadal timescale, which if confirmed would suggest the presence of torsional oscillations in the secular variation data\cite{Betal22}~\hspace{-1.5mm}. 
There are however still uncertainties on the exact depth at which this velocity is measured.}

We therefore suppose the spacing of the jets represents the spatial scale of the driving term of the oscillations.
The mean latitudes and the jet speeds have been given\editor{ 
\cite{TOLL17}~\hspace{-1.5mm}, and we use their 
OPAL2016 dataset
to construct Table~\ref{table:jets}.}
Since the 5\,$\mu$m observations \cite{AFetal19} extend reliably 
between latitude $\pm 40^\circ$, the extrema of the zonal wind speeds, together with the eastward jet speed ($-$ denotes westward) are shown, and the corresponding values of $s/R_\tx{J}$ are listed.
We then take the gap between two neighbouring peak latitudes\rev{, extending in $z$ down to \editor{the} top of the conductive region,}
 as an estimate of a half wavelength $\pi/k$ in $s$ (\editor{Methods}).
\editor{An alternative method 
is} to use the Lomb-Scargle technique applied to the zonal velocity data as a function of $s$.
\editor{Reassuringly, 
between latitudes $\pm 40^\circ$} we find a dominant maximum wavenumber of 13.0 in the north and 12.8 in the south, very similar in both hemispheres \editor{(Supplementary Figure~3).
This agrees with the wavenumbers predicted by the extrema method in Table~\ref{table:jets}.}

In Table~\ref{table:periods}, the periods found in the 5\,$\mu$m data\cite{AFetal19} are shown with \editor{the 
corresponding $s/R_\tx{J}$ values.}
The fourth column gives the Alfv\'en speed at these $s$-values; 
the fifth column gives the wavenumber based on the value in Table~\ref{table:jets} closest to the given latitude range. 
\editor{Then the half period, $T/2=\pi/U_\tx{A} k$, is listed in the final column.}
We give the half period based on the locally evaluated wavenumber, 
but also give in brackets the one 
based on the more global Lomb-Scargle approach. 
We marginally prefer the local
wavenumber, as the larger gap between the jets at latitudes $21^\circ$N and $28^\circ$N 
\editor{(also $24^\circ$S and $29^\circ$S)},
leading to the lower $k$ there, does seem to be a robust feature.

Torsional waves can only exist in an electrically conducting medium. 
Whilst the estimates of electrical conductivity\cite{FBLNBWR12} suggest that at equatorial latitudes between $\pm 16^\circ$ the currents are negligible there, the updated inference from gravity and magnetic Juno measurements\cite{GK21} indicated a transition at $\sim 0.972R_\tx{J}$, corresponding to $\sim \pm13.5^\circ$. 
We thus put a superscript~$^*$ in the fourth and sixth columns of Table~\ref{table:periods} where no torsional waves are, or could be, expected.   
The half period is found to be 
 \rev{4.6-4.7} years at the NTB (21-23$^\circ$ N), 
 \rev{close to} the 4.8 year period seen in the 5\,$\mu$m data. 
This is perhaps the most reliable of the computed torsional wave half periods, since the torsional wave cylinders at this latitude are deep enough that diffusion is only playing a minor role,
and yet far enough out for us to be confident that the potential field is a reasonable approximation to the true field.
The NTB disturbances are seen to break out at the same time in multiple disconnected regions at different longitudes\rev{\cite{AFetal18,AFetal19}}~\hspace{-1.5mm}. 
This is natural if they are triggered by a torsional wave, but \editor{hard to explain if they are triggered only by local convection}.
The half periods in 
\editor{the NTZ and the STB} are less than half the observed periods, but that could well be because the potential field approximation is not very good at depths below $0.9 R_\tx{J}$. 
Indeed, \editor{as} the magnetic coupling at $r=0.9R_\tx{J}$ is so strongly influenced by the \lq great blue spot', if the magnetic field lines associated with this feature spread out at depth as seems likely, they will no longer be primarily in the $s$-direction. 
This would reduce the magnetic coupling and hence lengthen the torsional \editor{wave 
period.} 
\editor{The NEB} half period does agree reasonably well with the observed period. 
This may be because the electrical conductivity falls off slightly less rapidly with height than current models suggest, or because the heat flux modulations there are coming from depth and so the half period in the equatorial regions comes from torsional oscillations at a smaller $s$-value than the nominal value in Table~\ref{table:periods}. 
Overall, the agreement between the observed and theoretically computed periods is remarkably good, especially where we would expect there to be good agreement.

\vspace{5mm}

\subsection*{\rev{Impacts of oscillations on the cloud-level convection}}

\rev{
We note that the half period is of interest because an interval identified in the brightness data
 is likely to depend on the square of the oscillating quantities
 just as the 11-year solar activity cycle arises from the 22-year magnetic oscillation.
Since we do not yet know the mechanism by which the oscillation causes the brightness fluctuations
 this is not altogether certain, and 
 it is possible to imagine mechanisms that could produce a full period cycle. 
Our mechanism of disruption of convection by the torsional wave shear would produce a half-period cycle.
The brightness fluctuations likely reflect changes in the convection, which could be affected by the large scale fluctuations in the zonal flow created by the torsional waves.}

\rev{
The effect of shear flow on \editor{convection} has been examined\editor{
\cite{JIN2022}~\hspace{-1.5mm}.
It was found} that the large scales of convection, which carry a significant amount of the heat, are disrupted by the shear flow velocity when its magnitude is comparable to the upwelling velocity of the convection. 
We do not know if \editor{this} extends to \editor{convection} in Jupiter. 
However, we apply this simple physical picture to explore when a torsional oscillation of order 1-10 cm/sec, consistent with numerical simulations\cite{HTJ19} and the secular variation data\cite{Betal22}~\hspace{-1.5mm}, can significantly affect the \editor{convective heat flux.}
A number of ways of estimating the convective velocity in a rotating system have been proposed \editor{(Methods and Supplementary Table~1)}. 
All give \editor{a rapid drop in the 
velocity with depth,}
because lower velocities there carry the same heat flux \editor{as} the density is higher. 
\editor{One formula\cite{SHOW11}}
gives 7\,cm/sec at 400\,km depth \editor{and} 
0.6\,cm/sec at 2500\,km depth. 
The Coriolis-Inertia-Archimedean theory\editor{\cite{AUR20} gives} slightly larger velocities, 20\,cm/sec at depth 400\,km and 4\,cm/sec at 2500\,km. 
These differences reflect the uncertainty\editor{,}
owing to our limited understanding of rotating convection\editor{.} 
Nevertheless, it seems likely that torsional oscillations of amplitude 1-10 cm/sec could affect the heat transport at depth, 
\editor{but not}
near the surface. 
However, a periodic blockage at depth could well lead to a modulation of the surface heat flux at the 5\% level. 
\editor{This could trigger convective storms} in the weather layer\editor{; this process needs} to be examined \editor{in} tropospheric cloud models\editor{\cite{HS01,SNetal11}~\hspace{-1.5mm}.}} 

\rev{
A further complication is the possible existence of stably stratified regions in Jupiter's interior\cite{DC19,GW21}~\hspace{-1.5mm}. 
If the buoyancy frequency is comparable with the rotation rate, as is generally believed for
\editor{planets,} then torsional oscillations are no longer $z$-independent as indicated in Figure~\ref{fig:torsional}\editor{, because the amplitude 
can vary in the stable region 
(Methods}). 
If there is a stable layer near the equatorial region of the torsional oscillations, \editor{the signal in the two
hemispheres could be different, possibly of the opposite phase
(\editor{Supplementary Figs.~2f-h})}. 
There is some evidence that the signal is generally weaker in the southern hemisphere, and 
is out of phase near latitudes 16$^\circ$, 22$^\circ$, and 33$^\circ$ in infrared observations\cite{AFetal19,OAFetal22}~\hspace{-1.5mm}.
The location of stably stratified zones is uncertain, but a zone out to $0.93R_\text{J}$ has been suggested\cite{DC19}~\hspace{-1.5mm}. 
If the torsional oscillation cylinder has a large region in the \editor{stable}
layer it might make small changes to the period, \editor{adding 
an uncertainty} to the expected periods \editor{at the higher latitudes.}
}

\vspace{5mm}

\subsection*{Brightness global map in the oscillation period range} 

\begin{figure}[bh]
  \includegraphics{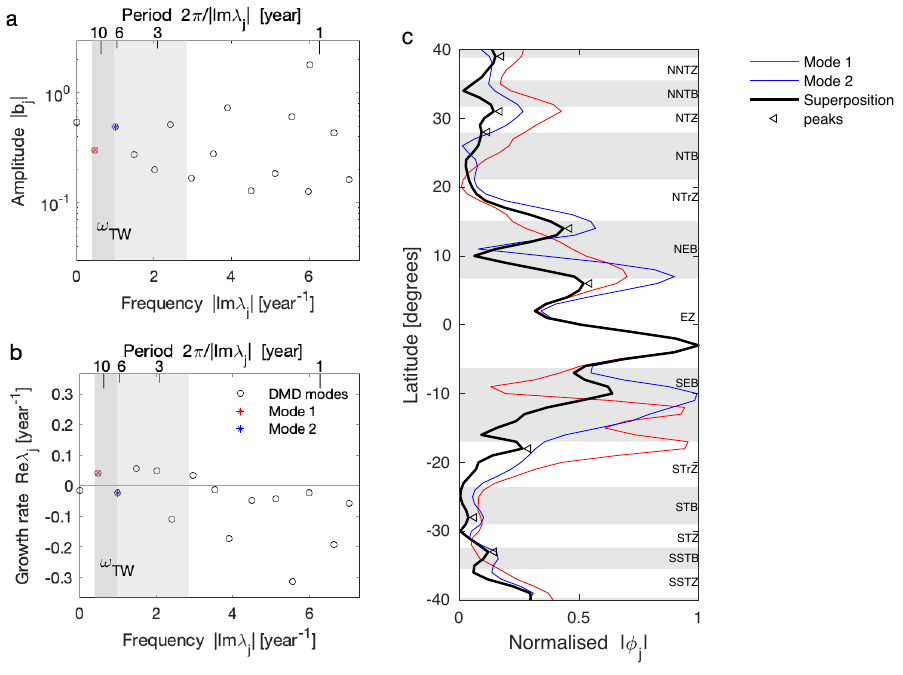}
\caption{(a) Spectrum and \editor{(b)} dispersion diagram of DMD modes of Jupiter's 5\,$\mu$m brightness\cite{AFetal19}~\hspace{-1.5mm}, $L'(\theta,t)$,
from March 2005 to July 2018 between latitudes $\pm 40^\circ$, 
 for coordinate delay $d = 1$ and no truncation in singular values. See details and their performance in Methods. 
 The shaded regions indicate the wave frequency ranges, $\omega_\tx{TW} = U_\tx{A}k$, as given in Table~\ref{table:periods}; the dark grey highlights the range predicted for latitudes between $\pm 24^\circ$ (i.e. rows 2-5 of Table~\ref{table:periods}).
 Red and blue asterisks indicate Modes 1 and 2, respectively, which are admitted in the wave window. 
 Mode 1 has half period\editor{, $\pi/\textrm{Im}\lambda_j$,} of 6.5 years and quality factor\editor{, $|\textrm{Im}\lambda_j/2\textrm{Re}\lambda_j|$,} of 5.8; 
 Mode 2 has half period of 3.1 years and quality factor of 23.
 \editor{(c)} Normalised, latitudinal profiles of Mode 1 (red curve), Mode 2 (blue), and their superposition (black)\editor{, i.e. $|b_\tx{Mode1}\;\phi_\tx{Mode1} (\theta) + b_\tx{Mode2}\;\phi_\tx{Mode2} (\theta)|$}. Triangles mark local peaks of the black curve, whose latitudes approximately match the jet ones.
 \editor{The jet latitudes and bands 
 are indicated in the shaded/white regions: for understanding their labels, see column 1 of Table~\ref{table:jets}.}} 
 \label{fig:brightness_dmd}
\end{figure}
\begin{figure}[bh]
  \includegraphics{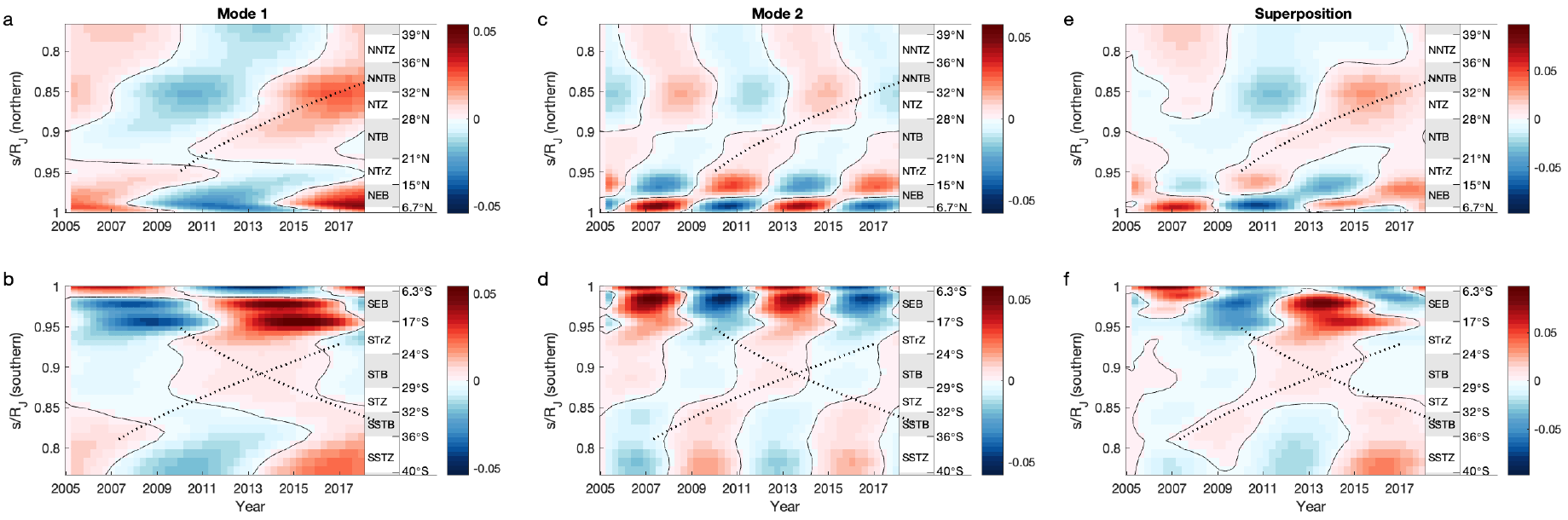}
\caption{\editor{Reconstructed spatiotemporal structures of the DMD Mode 1 (a-b), of the Mode 2 (c-d),
 and of their superposition (e-f) to reveal the
 wave motion.}
 Here the latitudinal profile is presented against the normalised cylindrical radius, $s/R_\tx{J}$, 
 in the northern \editor{(a, c, and e)} and southern \editor{(b, d, and f)} hemispheres. 
\editor{In each panel
 the jet latitudes and bands are presented on the right side: see column 1 of Table~\ref{table:jets} for labels.}
 Dotted curves indicate phase/ray paths of the torsional waves.
 Their slope, the predicted wave speed $U_\tx{A}$, explains the travelling nature of the infrared emission at $s/R_\tx{J} \lesssim 0.96$, i.e. in latitudes $\gtrsim 16^\circ$N/S.}
 \label{fig:brightness_dmd_s-t}
\end{figure}

\editor{Based on the expected wave speed, wavenumbers and periods,
we investigate spatiotemporal structures in the 5\,$\mu$m data\cite{AFetal19}~\hspace{-1.5mm}.
These might} correspond to travelling or standing waves in the $s$-direction:
both are possible for torsional waves\editor{.
}
%
%
\editor{
The global nature of the radiance signal is now filtered in terms of DMD, dynamic mode decomposition\cite{S10,RMBSH09,TRL14,KBBP16} (Methods).
}
%
Figures~\ref{fig:brightness_dmd}a and b display the dispersion and spectrum properties, respectively, of the \editor{resulting DMD modes.} 
Shaded regions indicate the wave frequency range, $\omega_\tx{TW}= U_\tx{A} k$, predicted above.
The spectrum is dominated by modes that have low frequencies and high quality factors\editor{. 
}
\rev{The wave window admits five modes, two of which are in the torsional oscillation frequency band,} one with 
 half period 
\editor{$\sim$6.5 years} 
 (referred to as Mode 1 hereafter; highlighted in red in the figures) and the other with half period
 $\sim$3.1 years 
 (Mode 2; \editor{in blue}), although the exact values vary by up to 4\% dependent on the chosen latitudinal/temporal ranges (\editor{Supplementary Table~2}).
\rev{Their latitudinal structures are exhibited in Figure~\ref{fig:brightness_dmd}c: we shall discuss details below.}
The reconstructed spatiotemporal structure takes the form of a travelling wave for Mode 1, but Mode 2 is more like a standing wave \editor{(Figures~\ref{fig:brightness_dmd_s-t}a-d)}. 
The superposition of the two Modes 
 is depicted in \editor{Figs.~\ref{fig:brightness_dmd_s-t}e-f} to reveal the wave nature migrating in $s$. 
\editor{In the figures,} dotted curves indicate arbitrary phase/ray paths of the predicted waves for guidance.
Their slopes, $U_\tx{A}$, appear to account for the brightness pattern, particularly in the northern hemisphere. 
Also, the retrieved spatiotemporal structure contains the bright points at mid latitudes in 2015-2017,
 as noticeable in \editor{the original dataset (Supplementary Fig.~4)},
 indicating those events were linked to the torsional wave.
This further suggests an equivalent event may occur in 2028-2030.

\rev{The signal's relation to the surface zonal wind structure, which we assumed for the period prediction,
 is examined in terms of the latitudinal structure\editor{.
 } 
In Figure~\ref{fig:brightness_dmd}c we demonstrate the normalised profile 
\editor{Mode 1, 
Mode 2, 
and  
their superposition.}
Here the jet-based belts and zones (Table~\ref{table:jets}) are indicated by shaded and white regions, respectively. 
Each Mode shows the peak at 3-12$^\circ$S, but also local ones at mid latitudes,
 such as 31$^\circ$N for both Modes and 14$^\circ$N for Mode 2:
 the former approximately matches the northern edge of 
 \editor{the NTZ} 
 while the latter matches the northern edge of 
 \editor{the NEB.}
The superposed signal in the black curve reveals the jet positions better.
Its peaks are found at latitudes including 31$^\circ$N, 28$^\circ$N, 14$^\circ$N, 18$^\circ$S, 28$^\circ$S, and 33$^\circ$S, 
and are indicated by triangles in the figure, showing a good correspondence with jets.
Their values are compared in Supplementary Table~3.
We conclude that the signal extracted from the brightness certainly illustrates wave-related \editor{lengthscales} as expected.}


\vspace{5mm}

\subsection*{Discussion} 

Despite the striking agreements shown above,
 \editor{we highlight} current uncertainties in the emission data.
First, the dataset was constructed from different instruments spanning several decades,
so the quality of the sets of infrared images cannot be compared straightforwardly\editor{,
but a technique like DMD can be beneficial.  
The above analysis} was capable of extracting signals with periods of several years existing in a limited timeseries,
 using only the images which were taken with modern, high-quality instruments. 
Secondly, 
there is potential bias because the 5\,$\mu$m imaging data cannot provide certainty over the vertical location of the aerosol thinning: it could be in the deeper clouds near 4 bars, or in the aerosols in the upper troposphere.
This could be rectified via spectroscopy over a broad wavelength range to determine exactly where the changes are occurring. 
This would allow further investigation of the hypothesis by means of dynamical simulations.

A remaining question is how exactly the torsional wave couples with the weather layer dynamics to generate the radiance variability. 
Deep dynamo simulations\editor{\cite{HTJ19} demonstrated that waves in the interior} 
could transmit towards the surface to modulate dynamical \editor{variables} at many latitudes,
and \editor{affect thermal properties near the equator.}
The tropospheric convection by itself may yield a quasi-periodic or intermittent cloud cycle at each site\cite{SNetal11}~\hspace{-1.5mm},
 but disturbances from below could further enforce such cycles in moist convection\rev{, through generating abrupt plumes; 
 we here discussed how the shear associated with a torsional oscillation could affect the convection.}
Whilst the \lq shallow' and \lq deep' models tend to be explored individually,
our findings push forward the need for further
 study of the coupling between the gas giant's atmospheric dynamics and the 
 internal dynamics, which is quite distinct from the dynamical meteorology of the Earth.

\rev{The cloud-top zonal wind has been seen to vary on timescales of 6-14 years
 with magnitude smaller than 10\% of the $\sim$150\,m/sec stable jets\cite{TOLL17,Wetal20}~\hspace{-1.5mm}. 
The velocity measurement was based on horizontal correlations in multiple images at visible wavelength.
Those snapshots since late 2008 were used to derive the velocity profile in 9 different epochs and to find its variation, significantly at 21$^\circ$N (NTB) and 8.8-13$^\circ$S (SEB)\cite{Wetal20}~\hspace{-1.5mm}\editor{.}
\rev{The amplitude of the deep torsional wave is too small to be detected currently in comparison with the violent convection and zonal flow near the surface. 
It is only deep down, where the convection is relatively slow and the zonal flows are expected to be much steadier, that the torsional oscillation can have a significant dynamical influence by modulating the heat transport}.  
Currently the \editor{5$\mu$m} measurement as discussed in this study achieved a much larger set of data, more than 23000 snapshots during the same epochs.
This enabled us to analyse the time-varying component more precisely. 
Future updates of the cloud-top zonal wind will help to test these findings derived from the infrared observations.}

\editor{The} axisymmetric magnetic waves may trigger fluctuations near the surface\editor{, 
 but} they could also lead to variations in the planet's rotation rate and its magnetic secular variation\editor{\cite{HTJ19}
 as they do for the Earth\cite{GJCF10}~\hspace{-1.5mm}.}
\makeatletter
The rotation rate is determined from the periodic radio emission of the gas giant (the System \@Roman{3}).
It has been suggested that some significant variations have already been detected\cite{HCR96}~\hspace{-1.5mm}.
\makeatother  
The variation of the intrinsic magnetic field is going to be better resolved by Juno magnetometer measurements\cite{Betal22}~\hspace{-1.5mm}.
\rev{Given a steady field magnitude $|B| \sim$ 3\,mT, a flow fluctuation $|u'| \sim$ 1\,cm/sec, a typical lengthscale $D \sim$ 0.2$R_\tx{J}$, and a wave period $T \sim$ 10\,years, then a magnetic fluctuation near the conductive region is found to be  $|B|\,|u'| T/2\pi D \sim$ 0.1\,mT and a secular variation as $|B|\,|u'|/D \sim$ 0.07\,mT/yr\editor{,}
comparable with the observed variation \cite{Betal22}~\hspace{-1.5mm}.}
\rev{We note that, when there are layerings in the density\cite{Metal18,GW21}~\hspace{-1.5mm}, magnetic signatures can be significantly weaker near the equator and be equatorially asymmetric. 
The presence of a stratified layer prevents torsional oscillations, excited near the surface in one hemisphere, spreading over $z$ in the interior (\editor{Supplementary Fig.~2}).
This is likely compatible with the current observation, albeit it has not excluded the possibility of rapid variations other than those measured\cite{Betal22}~\hspace{-1.5mm}.}

An important consequence of wave exploration is the possibility of seismology in gaseous planets.
Analogous to the successful technique of helioseismology,
 advances in dioseismology have been made to determine the structures and properties of acoustic waves therein\cite{GSGGJ11}~\hspace{-1.5mm}; 
 however, the rapid rotation hinders a simple translation of helioseismological techniques. 
Other wave classes have been proposed to probe the existence of a solid core\cite{G18}~\hspace{-1.5mm},
 which may well be diluted\cite{WHMetal17,S20}~\hspace{-1.5mm}, 
 and  
 \rev{of the stratified layers}
 as characterised by internal gravity \editor{waves\cite{PBetal20}~\hspace{-1.5mm}.
Our} magnetohydrodynamic waves are capable of scanning the strongly electrically conducting region, and hence may put similar constraints \editor{on 
the magnetic field within} the deep interior.
Jupiter's conducting region is thought to occupy the major part of its volume, and
 to generate the strongest planetary magnetic field in our Solar System. 
Observations of magnetic waves could provide us with strong constraints
 on the internal structure, the mechanisms of formation of the planet, and how the dynamo generates Jupiter's  magnetic field.

\clearpage
\section*{Methods} 

\begin{description}
\item[Torsional wave/oscillation equation]:
  Dissipationless torsional waves for electrically-conducting, anelastic fluids are given by
 the second-order differential equation for the fluctuating part of the zonal velocity, 
 $u'_\phi (s,\phi,z,t)$ \editor{with $t$ being time}: 
\begin{equation}
 \frac{\partial^2}{\partial t^2} \frac{\langle \overline{u'_\phi} \rangle}{s} 
 - \frac{1}{s^3 h \langle \overline{\rho} \rangle } \frac{\partial}{\partial s}
     \left( 
       s^3 h \langle \overline{\rho} \rangle
       U_\tx{A}^2
       \frac{\partial}{\partial s} \frac{\langle \overline{u'_\phi} \rangle}{s}
     \right) 
 = \frac{\partial }{\partial t} \frac{F_\text{R} + F_\text{LD}}{s \langle \overline{\rho} \rangle }
  \label{eq:TW}
\end{equation}
\cite{Bra70,RA12,HTJ19}~\hspace{-1.5mm}. 
Here $\overline{f}$ and $\langle f \rangle$ denote the $\phi$-average (i.e. the axisymmetric part)
 and the $z$-average from $z_{+}$ to $z_{-}$, respectively, for an arbitrary function $f$.
Also $h = z_{+} - z_{-}$ is the height of the cylinder of radius $s$ along the $z$ axis:
 in this study we suppose $z_\pm = \pm \sqrt{R_\tx{J}^2 - s^2}$ with $R_\tx{J}$ being the mean radius of Jupiter, $R_\tx{J}=69894$\,km\cite{FBLNBWR12}~\hspace{-1.5mm}.
The Reynolds force $F_\text{R}$ is defined by 
 \begin{eqnarray}
 F_\tx{R} &=&
   - \frac{1}{s^2 h}\frac{\partial}{\partial s}
      s^2 h \left\langle \overline{\rho} \; \overline{ u_s u_\phi } \right\rangle
   - \frac{1}{h} \left[
                  \frac{s}{z} \overline{\rho} \; \overline{u_s u_\phi}
                  + \overline{\rho} \; \overline{u_z u_\phi}
                 \right]_{z_{-}}^{z_{+}}  \;, 
 \end{eqnarray}
whilst $F_\text{LD}$ denotes the Lorentz force, excluding the restoring part for the wave.
Eq.~(\ref{eq:TW}) is reduced to a wave equation when the right-hand side of the equation is omitted. 
Each term on the right-hand side may act as a forcing on the torsional wave.
In Jupiter's non-conducting envelope,
 zonal flows $u_\phi$ are known to be significant so that the $F_\text{R}$ term  provides the main driving\rev{\cite{HTJ19}}~\hspace{-1.5mm}.

 \vspace{5mm}
 
 \item[Calculation of $U_\text{A}$]:  
We require the calculation of the Alfv\'en speed
$U_\tx{A} = (\langle \overline{B_s^2} \rangle/\rev{\mu_0} \langle \overline{\rho} \rangle )^{1/2}$.
\rev{The density profile is taken from \editor{the model of French {\it et al.}\cite{FBLNBWR12}~\hspace{-1.5mm},} which was derived from gravity measurements and which should be sufficiently accurate for our purposes.
 We then compute the cylindrically averaged density $\langle \overline{\rho(s)} \rangle$}
 using their Table~1\cite{FBLNBWR12}~\hspace{-1.5mm},  
  which is given up to the radius $r_\text{max}=0.994 R_\tx{J}$.
So we take the $z$-average over $-z_\text{max} < z < z_\text{max}$ where $z_\text{max} = \sqrt{r_\text{max}^2-s^2}$, the density further out being negligible.

The field model \rev{JRM33 of \editor{Connerney {\it et al.}\cite{Cetal21}~\hspace{-1.5mm},} as well as the early one JRM09\cite{Cetal18}~\hspace{-1.5mm},}
  gives the Gauss coefficients of the magnetic potential $V$, where $\mib{B} = - \nabla V$ and
\begin{equation}
 V(r,\theta,\phi)
 = r_\tx{eq} \sum_{n=1}^\infty \sum_{m=0}^n \left(\frac{r_\tx{eq}}{r} \right)^{n+1}
     P_n^m(\cos\theta) [g_n^m \cos m\phi + h_n^m \sin m\phi]  \; .  
 \label{eq:potential}
\end{equation}
Here $r_\tx{eq}=71492$\,km is the equatorial radius of Jupiter.
The Gauss coefficients $g_n^m$ and $h_n^m$ refer to Schmidt normalised spherical harmonics $P_n^m$.
The field component $B_s$ is given in spherical polar coordinates $(r,\theta,\phi)$, where $z = r\cos{\theta}$ and $s = r \sin{\theta}$, by 
\begin{equation}
 B_s = - \frac{\partial V}{\partial s}
     = - \sin \theta \frac{\partial V}{\partial r} - \frac{\cos \theta}{r} \frac{\partial V}{\partial \theta} \; .
  \label{eq:potential_Bs}
\end{equation}
Using (\ref{eq:potential}) and the recurrence relation for the Schmidt normalised derivative,
\begin{equation}
 (1-x^2) \frac{d}{dx} P_n^m (x)= (n+1) x P_n^m(x) - \left( (n+1)^2 - m^2 \right)^{1/2} P_{n+1}^m (x) \; ,
\end{equation}
where $x=\cos \theta$, we can rewrite (\ref{eq:potential_Bs}) as 
\begin{equation}
 B_s =  \sum_{n=1}^\infty \sum_{m=0}^n \left(\frac{r_\tx{eq}}{r} \right)^{n+2} [g_n^m \cos m\phi + h_n^m \sin m\phi] \left( \frac{n+1}{\sin \theta} P_n^m(x)
  - \frac{\cos \theta}{\sin \theta} [(n+1)^2 - m^2]^{1/2} P_{n+1}^m (x) \right) \; .
\end{equation}
We compute this
 using the coefficients \rev{up to degree $n = 18$ listed in the Supporting Information Table~S1 of Connerney {\it et al.}\cite{Cetal21}}~\hspace{-1.5mm}.
The Schmidt normalised associated Legendre function is evaluated by the Matlab function \lq legendre'
 with the option \lq sch'. 
Outside $r = r_\text{diff}=0.96R_J$ magnetic diffusion will greatly reduce the magnetic restoring force, so to   
take diffusion into account in a simple way, we integrate $B_s^2 (s,\phi,z)$ only over $\phi$ and $-z_\text{diff} < z < z_\text{diff}$ where $z_\text{diff} = {\sqrt{r_\text{diff}^2 - s^2}}$, thus omitting the restoring force where diffusion dominates, but we divide by the full cylinder area $4 \pi s z_\text{max}$ to
obtain $\langle \overline{B_s^2} \rangle/\mu_0$ \rev{where $\mu_0 = 4\pi \times 10^{-7}$ H\;m$^{-1}$}.

\rev{In Supplementary Figure~1, the wave speed $U_\tx{A}$ is computed using both the new JRM33 model\cite{Cetal21}~\hspace{-1.5mm} and the older JRM09\cite{Cetal18} model,
which was based on the measurements during the early 9 orbits in 2016-17; cf. 33 orbits until April 2021 for JRM33.
Since the JRM09 model only goes up to $n=10$ while JRM33 goes up to $n=18$, we show the results from JRM33 using both the full $n=18$ and JRM33 truncated at $n=10$. 
The $n=10$ JRM09 and JRM33 models (blue and black curves) are similar, despite the differing measurement intervals, but using the full $n=18$ harmonics increases the wave speed, particularly at smaller radii.
The speeds and the corresponding wave periods at the selected latitudinal regions are depicted in Supplementary Table~4 to be compared with Table~\ref{table:periods}.
The period predictions with the high resolution JRM33 model fit best with the intervals identified in the infrared observation.
We hence present this case in the main text, Table~\ref{table:periods}, and all other figures presented in the manuscript.
}

\rev{Here the Lundquist number\editor{,} representing the ratio of the wave speed to the diffusive one,
 is calculated as $(\overline{B_s^2}/\mu_0\overline{\rho})^{1/2} D/\eta$ at $z = 0$
 (where $\eta$ and $D$ are the magnetic diffusivity\cite{FBLNBWR12} and a typical lengthscale, respectively)
 to show its abrupt change with radius, \editor{varying from}
 the order of $10^6$ to $10^1$ (\editor{Supplementary ~Fig.~1b}).}

 \vspace{5mm}

 \item[Calculation of $k$]:   
   We calculate the wavenumber $k$ at a given cylindrical radius $s$ in the following two ways.

   The first method relies on the gap widths of the jets visible at the gas giant's surface. 
   We calculate the planetocentric latitudes of the jets' extrema using the dataset \editor{Outer Planetary Atmosphere Legacy (OPAL)} 2016 in Tollefson {\it et al.}\cite{TOLL17}~\hspace{-1.5mm}, 
   convert them into the cylindrical coordinates, and take the interjet spacings as a half wavelength $\pi/k$. 
   We choose two neighbouring jets so that their planetocentric latitudes bound the region of interest in Table~\ref{table:periods}.
   For example, the gap between the $\sim 21^\circ$N jet and the $\sim 29^\circ$N jet
   is assumed to give the local wavelength for the Table~\ref{table:periods} region of interest in the NTB, 21-23$^\circ$N:
   similarly, the gap between the $\sim 6.3^\circ$S and $\sim 17^\circ$S jets is used for the region in the SEB, 12-15$^\circ$S.

   The alternative approach is to perform a Lomb-Scargle analysis of the jet speed profile OPAL 2016, using only the data between $\pm 40^\circ$ in each hemisphere.
   The Lomb-Scargle periodogram fits an irregularly sampled data series onto
   a sinusoidal function for a given frequency by least squares\cite{S82}~\hspace{-1.5mm}.
   We use the Matlab function `plomb' with the option `normalized' and a wavenumber spacing of 0.1 
   to seek peaks located in the relevant wavenumber windows: $5 \le k R_\tx{J}/2\pi \le 20$ for both the  northern and southern jets. 
  The spatial spectra are displayed in Supplementary Fig.~3
  and have a single dominant peak which matches well with our expectations from the first method.
  The other datasets given by \cite{TOLL17} give a similar dominant peak, so this approach appears to be robust.
  We don't use the lesser peaks in Supplementary Fig.~3 as these are not robust between the different datasets. 

  \vspace{5mm}

 \item[\rev{Forward simulations of torsional oscillations excited by zonal wind variation}]\rev{: To examine how the near-surface zonal flow forces the deep oscillations and what their magnetic signature looks like, we consider a linearised, anelastic magnetohydrodynamic model. }
 
 \rev{We take Cartesian coordinates with $x$ aligned with the cylindrical $s$ coordinate, $y$ eastward (in the $\phi$ direction) and $z$ parallel to the rotation axis $\mib{\Omega}$. 
 For simplicity all variables are assumed independent of $y$, such that 
 momentum $\rho \mib{u}=\nabla \times \psi(x,z,t) \hat{\mib{e}}_y$,
 magnetic field $\mib{B}= B_0 \hat{\mib{e}}_x + \nabla \times A(x,z,t) \hat{\mib{e}}_y + B_y(x,z,t) \hat{\mib{e}}_y$,
 and entropy $S_0 (z) + S(x,z,t)$.
 Here $B_0$ is constant and $\hat{\quad}$ denotes a unit vector.
 Given a zonal flow forcing $f(x,z,t)$, we investigate the time evolution of the $y$-components of the velocity
 \begin{equation}
    \frac{\partial u_y}{\partial t} 
    = -2\Omega u_x + \frac{B_0}{\mu_0 \rho} \frac{\partial B_y}{\partial x} 
      + \frac{\mu}{\rho} \left( \frac{\partial^2 u_y}{\partial x^2} 
                   + \frac{\partial^2 u_y}{\partial z^2} \right) + f(x,z,t),
 \end{equation}
 with $\mu$ being the dynamic viscosity, assumed constant, 
 of the vorticity
  \begin{equation}
    \frac{\partial \omega_y}{\partial t} = 2\Omega \frac{\partial u_y}{\partial z}
     + \frac{B_0}{\mu_0 \rho} \frac{\partial j_y}{\partial x} 
     - \frac{g_z}{c_p} \frac{\partial S}{\partial x}
     + \frac{\mu}{\rho} \frac{\partial^2 \omega_y}{\partial x^2}
     + \frac{\partial}{\partial z}\frac{\mu}{\rho} \frac{\partial\omega_y}{\partial z},
 \end{equation}
 with $c_p$ being the specific heat at constant pressure, 
 of the magnetic field
 \begin{equation}
   \frac{\partial B_y}{\partial t}
   = B_0 \frac{\partial u_y}{\partial x}
    + \eta \frac{\partial^2 B_y}{\partial x^2}
    + \frac{\partial}{\partial z} \eta \frac{\partial B_y}{\partial z},
 \end{equation}
 with $\eta = 1/\mu_0 \sigma$ and $\sigma$ being the magnetic diffusivity and the electrical conductivity, 
 and of the electric current 
 \begin{equation}
   \frac{\partial j_y}{\partial t}
   = \frac{B_0}{\mu_0} \left( \frac{\partial \omega_y}{\partial x}
    + \frac{\partial}{\partial z} \frac{1}{\rho^2} \frac{d\rho}{dz} \frac{\partial \psi}{\partial x} \right)
    + \eta \frac{\partial^2 j_y}{\partial x^2}
    + \frac{\partial^2}{\partial z^2} \eta j_y \;,
 \end{equation}
 and the entropy
  \begin{equation}
   \frac{\partial S}{\partial t} 
   = - \frac{1}{\rho} \frac{dS_0}{dz} \frac{\partial \psi}{\partial x} 
    + \kappa_s \left( \frac{\partial^2 S}{\partial x^2} 
                   + \frac{\partial^2 S}{\partial z^2} \right),
 \end{equation}
 with $\kappa_s$ being the entropy diffusion coefficient, assumed constant.
Here the vorticity and the current are given with the streamfunction and potential, respectively, as 
\begin{equation}
 \omega_y = -\frac{1}{\rho} \frac{\partial^2 \psi}{\partial x^2}
           - \frac{\partial}{\partial z} \frac{1}{\rho} \frac{\partial \psi}{\partial z}
 \quad \text{and} \quad 
 j_y = - \frac{1}{\mu_0} \left( \frac{\partial^2 A}{\partial x^2} 
                   + \frac{\partial^2 A}{\partial z^2} \right) \;.
\end{equation}
Of particular interest is the evolution of a mode $\exp{\mathrm{i} k_x x}$ with zonal flow $u_y$ being modulated near the surface in the northern hemisphere. 
We then suppose the forcing is given as
\editor{$f = (\exp{\mathrm{i} k_x x} \exp{\mathrm{i} \omega_0 t}) F(z)$ where $\omega_0$ is a forcing frequency:
 the $z$-profile $F(z)$ is demonstrated in Supplementary Figure~2a.}
The density profile $\rho$ is assumed to be a polytropic reference state of index $1$, and the conductivity $\sigma$ to represent a transition between the molecular and metallic envelopes (\editor{Supplementary Figs.~2b-c}).
A stably stratified layer can be modelled with a given entropy profile,
 \editor{the gradient of which, multiplied by gravity, is shown in Supplementary Fig.~2f}. 
The gravity profile $g_z \propto z/z_0$ is assumed for the buoyancy term.
We scale length so that wavenumber $k_x$ should become unity, so a unit length of $\sim 1100$km.
\editor{We then suppose a column running at latitudes $\sim 21^\circ$N/S, from $-$25000\,km to $+$25000\,km in $z$: our dimensionless $z$ may range from $-$23 to $+$23.} 
Time is scaled so that the torsional wave frequency $k_x U_\tx{A}$ should be unity,
 yielding that our time unit here is $\sim$1.5 years.  
The location and strength of the stratified layer are very uncertain; adopting the model of \cite{GW21}
\editor{we suppose it starts at a depth near 4300\,km below the surface (see Supplementary Fig.~2f),
 and we set the buoyancy frequency to be $1.5|\mib{\Omega}|$.}
We solve the 1d forward problems by using the Matlab function pdepe with 181 gridpoints in $z$. 
}

\rev{
Supplementary Figures~2d and e depict the evolution of the zonal flow perturbation $U(z,t)$ and the magnetic one $B(z,t)$, respectively, when stably stratified layers are absent.
Given the forcing of $\omega_0=0.88$ at $t=0$, the flow modulation instantly spreads from the northern surface, $z=+20$, to the southern hemisphere, $z<0$, and excites torsional oscillations of frequency $\sim 1$ that are symmetric with respect to the equator, $z=0$ (figure d). This gives rise to an equatorially-symmetric variation in the magnetic component (figure e).}

\rev{
A case in the presence of a stratified layer is illustrated in \editor{Supplementary Figs.~2g and h}. 
The identical forcing leads to torsional oscillations being excited in the northern hemisphere down to the stratified layer, $5 \lesssim z$ (figure g). This leads to a magnetic signal that is remarkable in the northern hemisphere but much less clear for $z\lesssim 2.5$, i.e. around the equator and in the southern hemisphere.
This demonstrates that the deep magnetic oscillations may be excited from a near-surface wind modulation, regardless of the stably stratified layer, but that the presence of the layer would naturally give an asymmetry in the magnetic observation.
}

 \vspace{5mm}

\item[\rev{Convective velocity estimates and the effect of shear}]\rev{:
Estimates of the convective velocity as a function of depth are based on scaling laws backed up by laboratory experiments. It is not possible to reproduce conditions in planetary interiors conveniently, so there is considerable uncertainty in these results. 
Showman {\it et al.}\cite{SHOW11} suggested
\begin{equation}
 U_\text{Show}= \left( \frac{g \alpha F}{\rho c_p \Omega} \right)^{1/2}, \label{eq:showman}
\end{equation}
where $g$ is local gravity, $\alpha$ is the coefficient of thermal expansion, $F$ is the heat flux coming out of Jupiter ($\sim 5.4$ Wm$^{-2}$), $\rho$ is the density, $c_p$ is the specific heat at constant pressure and $\Omega$ is the rotation rate. 
Estimates of the thermodynamic quantities are available\cite{FBLNBWR12}~\hspace{-1.5mm}. 
Rapidly rotating convection often takes the form of tall thin columns aligned with the rotation axis, but (\ref{eq:showman}) takes no account of this. 
The Coriolis-Inertial-Archimedean (CIA) theory uses a three-term balance in the equation of motion (see e.g.~\cite{AUR20}), taking the width of the rolls as $\ell$ and their extent parallel to rotation as the pressure scale height $H$. 
$\ell$ is eliminated in the theory to give     
\begin{equation}
 U_\text{CIA}= \left( \frac{g \alpha F}{2 \rho c_p \Omega} \right)^{2/5} (2 H \Omega)^{1/5} \;.   
 \label{eq:cia}
\end{equation}
Supplementary Table~1 compares the estimates at different depths.
The Showman {\it et al.} formula gives a lower estimate for the velocity. 
In the CIA theory the roll width $\ell$, which is the Rhines length-scale\editor{,} is about 16\,km, much smaller than the roll height $H$. 
The CIA regime is hard to reach in experiments, and it is possible that turbulence limits the ratio $H/\ell$, so the best estimate below $1500$\,km depth probably lies somewhere between the CIA and the Showman {\it et al.} formulae, a few cm/sec. 
}

\rev{
The shear velocity from the torsional oscillation is likely also a few cm/sec at maximum amplitude, as the secular variation data is currently our most reliable guide, though further data from Juno will hopefully increase its reliability. 
This shear occurs over an $s$-range of about 2700\,km, the half-wavelength of the torsional oscillation in the $s$-direction. 
Simulations such as Jin {\it et al.}\cite{JIN2022} suggest that a shear of the same magnitude as the convective velocity, the situation we have here, can significantly affect the heat flux. 
We therefore consider it plausible that fluctuations of the order of 5\% could be responsible for the 5\,$\mu$m signal. 
There are, however, still important unknowns here. 
We know little about how shear affects rapidly rotating convection, as opposed to non-rotating convection. 
There is also uncertainty about how a modulation of the heat flux at a depth of 1000\,km or deeper is transmitted to the surface.     
}

  \vspace{5mm}

\item[Analysis and dynamic mode decomposition of brightness data]:
\editor{
We reanalyse the zonally-averaged 5\,$\mu$m brightness, $L (\rev{\theta},t)$, 
of Antu\~{n}ano {\it et al.}\cite{AFetal19}~\hspace{-1.5mm},
where $\rev{\theta}$ and $t$ are colatitude and time, respectively.
%
%
We separate 
$L (\theta,t)$} into the temporal mean at a specific latitude, $\widetilde{L}(\theta)$,
 and the anomaly, $L' (\theta,t) = L (\theta,t) - \widetilde{L}(\theta)$.
Here the unevenly sampled dataset is interpolated by cubic polynomials and smoothed by running averages with a window size of 100 days, 
into bins of size $\Delta \theta = 1^\circ$ and $\Delta t = 1$ day. 
The anomaly $L'$ is displayed in Supplementary Figure~4. 
The latitudinal profiles are presented in terms of the normalised cylindrical radius,
 $s/R_\tx{J} = \sin{\theta}$, in each hemisphere separately,
 assuming that the sources of the 5\,$\mu$m thermal emission and its variations
 are located at the 1-10 bar level,
 i.e. no smaller than 99.8\% of the gas planet's nominal radius $R_\tx{J}$.
The figure confirms that the
 prominent events occur recurrently at low latitudes $\lesssim$ 18$^\circ$ where $s/R_\tx{J} \gtrsim 0.95$.
A pair of events can be seen during 2015-2017 at $0.75 \lesssim s/R_\tx{J} \lesssim 0.85$,
 i.e. latitudes $30$-$40^\circ$.
\editor{Convective processes within the troposphere alone would not account for the global outbreaks that span large latitude ranges, and also large longitude ranges\cite{AFetal19}~\hspace{-1.5mm}. }

\editor{
We now perform DMD of the 
brightness data.
This method is often used to diagnose the fluid dynamics behind experimental data,
 by approximating the data as a superposition of modes of the form 
 $\sum_{j=1}^r b_j \phi_j (\theta) \exp{(\lambda_j t)}$ 
 where \rev{$b_j$ is real and $\lambda_j$ and $\phi_j$ are complex. 
 Here} $\textrm{Im}\{\lambda_j\}$, $\textrm{Re}\{\lambda_j\}$, $b_j$, and $\phi_j (\theta)$ denote
 the frequency, growth rate, amplitude, and latitudinal profile of the $j$-th DMD mode, respectively. 
The technique has been utilised for a wide variety of application areas
 including neuroscience, epidemiology, and traffic dynamics
 (e.g. \cite{KBBP16}, \cite{AM20} and references therein).
Here we shall apply the technique to observational data in astrophysics;
 previously magnetohydrodynamic oscillations and waves were distinguished in simulation data, 
 and compared well with the theoretically expected frequencies\cite{HTT20}~\hspace{-1.5mm}.
We implement the standard DMD of shift-stacking brightness data $L'$: 
 the methodology can be regarded as a Hankel DMD\cite{AM20}~\hspace{-1.5mm}.
}

\editor{
Our} overall dataset has 141 gridpoints in latitude and 123 snapshots, sampled every 100 days. 
To focus on some notable features (\editor{Supplementary Fig.~4}) but also to make the data quality comparable, 
 we here examine the data at 81 gridpoints ($=M$) between $\pm 40^\circ$ and for 49 snapshots ($=N$) 
 \editor{from March 2005 to July 2018}.
They correspond to the eras of four modern instruments: 
 MIRSI (January 2005 - April 2006), NSFCam2 (April 2006 - October 2008),
 SpeX (July 2009 - August 2017), and TEXES (December 2014 - July 2018). 
\editor{An overview of these instruments and measurements is given by \cite{AFetal19}~\hspace{-1.5mm}. }
Below 
we shall investigate the influence of the chosen latitudinal/temporal window on the performance of DMD and resultant Modes.


As torsional waves may have the form of standing waves,
 we exploit delay coordinates to stack time shifted copies of the data, 
 following the standard DMD algorithm.
For details of the methods we refer to \cite{KBBP16} and \cite{AM20}~\hspace{-1.5mm}. 
The mean-extracted data, ${L'}$, is represented as a matrix
 $X = [\mib{x}_1 \; ... \; \mib{x}_N ] \in \mathbb{R}^{M\times N}$ (where $\mib{x}_k \in \mathbb{R}^{M}$ with $k=1, ..., N$)
 and further set to form an augmented matrix, $X_\tx{aug} \in \mathbb{R}^{M(d+1)\times (N-d)}$,
 given the number $d$ of delays used. 
 Then singular value decomposition of $X_{\tx{aug},1} = [\mib{x}_1 \; ... \; \mib{x}_{N-1} ]$ yields $X_{\tx{aug},1} = U \Sigma V^\tx{T}$,
 where $\Sigma$ is an $r_\tx{f}\times r_\tx{f}$ diagonal matrix
 with diagonal entries $\sigma_j \ge 0$ ordered as $\sigma_1 \ge \sigma_2 \ge  ...$
 with rank $r_\tx{f}$.
 With $r$ modes (out of $r_\tx{f}$) being kept,
 the eigenvalue problem of the optimal matrix $A = U^\tx{T} X_{\tx{aug},2} V \Sigma^{-1}$ (where $X_{\tx{aug},2} = [\mib{x}_2 \; ... \; \mib{x}_N ]$) is solved
 to give DMD eigenvalues $\mu_j$ and eigenfunctions $\Phi = [\mib{\phi}_1 \; ... \; \mib{\phi}_j \; ... \; \mib{\phi}_r ]$.
 The frequency and growth rate of each mode is determined as $\lambda_j = \ln{\mu_j}/\Delta t$.
 DMD of the mean-extracted data must not yield zero eigenvalues, $\mu_j = 0$: so, 
 as guided by \cite{AM20}~\hspace{-1.5mm},
 we utilise this to validate the analysis and select the parameter $d$.
 Another parameter is the choice of truncation $r$ to form the optimal matrix $A$.

The amplitude may be computed from the initial data, $\mib{x}_1$, as
 $\mib{b}_\text{I} = \Phi^\dag \mib{x}_1 {2/\sqrt{1+d}}$ with $^\dag$ being the pseudoinverse matrix,
 or from the coefficients of a proper orthogonal decomposition,
 $\Sigma V^\tx{T}$, as 
 $\mib{b}_\text{P} =  \Phi^{-1} (\Sigma V^\tx{T}) {2/\sqrt{1+d}}$\cite{JSN14}~\hspace{-1.5mm}.
The performance of the decomposition and power estimation is measured in terms of 
 errors between the observables, $X$, and the reconstructed data, $X_\tx{rec}$: 
 the root mean square error, RMSE $= ||X - X_\tx{rec}||_\tx{F}/NM$\cite{AM20}~\hspace{-1.5mm}.

Supplementary Table~2 summarises the performance for different sets of the chosen latitudinal ranges, duration, coordinate delay $d$, and mode truncation $r$. 
Given the dataset betwen latitudes $\pm 40^\circ$ since March 2005, the use of delay coordinate $d=1$ confirms the absence of the spurious eigenvalue $\mu_j = 0$ and so the decomposition is valid: 
Supplementary Fig~5a shows all the $\mu_j$ for $d=1$.
In contrast a delay of $d=2$ likely suffers from the issue.
We therefore choose the minimal value, $d=1$, for the analysis we present.
This yields the rank $r_\tx{f} = 47$, and so there are 47 DMD modes when the all singular values are retained. 
Both amplitudes $\mib{b}_\text{I}$ and $\mib{b}_\text{P}$ lead to a small RMSE, implying that the data $X$ is reasonably reconstructed through the DMD.
As the formula $\mib{b}_\text{P}$ gives the lesser error, 
 we adopt $\mib{b}_\text{P}$ for the presented diagrams.

Next we examine the impact of the truncation $r$. 
The whole singular values $\sigma_j$ for the case are shown in Supplementary Fig.~5b. 
The figure shows a gap at $j \sim 40$. 
Thresholds to retain 99\% and 90\% of the cumulative energy imply
 a truncation at $j = 39$ (dotted vertical line) and $21$ (dashed-dotted), respectively;
 the optimal hard threshold for unknown, random noise\cite{GD14,KBBP16} gives $j = 15$ (dashed lines). 
Keeping 39 modes (corresponding to the 99\% threshold)
 ensures the two Modes of half period $T_j/2 = \pi/\text{Im}\lambda_j \sim$6.5 years and $\sim$3.2 years, 
 whilst their growth rates, $\text{Re}\lambda_j$, and so quality factors, $Q_j = |\text{Im}\lambda_j /2\text{Re}\lambda_j|$, vary.
We do not find either when retaining 21 modes only (the 90\% threshold) or fewer modes. 
This suggests that the current dataset requires a large number of modes to converge the eigenvalues.
Since the computations are quite cheap, 
 we decided to retain the whole set of singular values in the current work.

Finally we check the sensitivity to the latitude-time window.
When the temporal window is taken from Nov 2003 (i.e. the number of snapshots $N=54$), the same procedure gives the two Modes of half period $\sim$ 6.2 years and $\sim$ 3.1 years, which is within 4\% error of the above case.
The equivalent Modes are found for a wider latitudinal window spanning over $\pm 70^\circ$ ($M =141$).
Interestingly we find them even when the dataset is restricted to latitudes 16-40$^\circ$ ($M =50$), where the equatorial region is excluded from the analysis. 
Their quality factors may vary, however they are fairly high to ensure the wave feature. 
As a whole the spatiotemporal structure of the superposition does not significantly change from \editor{Figs.~\ref{fig:brightness_dmd_s-t}e-f}.
We thus conclude the DMD successfully extracts the several-year signals hidden in the original brightness dataset.

\rev{To characterise the spatial structure $|\phi_j (\theta)|$, as shown in Fig.~\ref{fig:brightness_dmd}c, we use the Matlab routine 'findpeaks' with no options. 
This endorses its peaks and minima at latitudes. The results are listed in Supplementary Table~3 and are compared to the jet latitudes.}
%
%
\rev{\editor{We here 
remark the spatial structures of Modes 1 and 2 \editor{exhibits} two and three minimal points, respectively, in either hemisphere. 
They could imply the number of nodes that oscillations have, suggesting Mode 2 is a higher eigenmode by one degree than Mode 1.
This is consistent with their periods: $T_j/2$ for Mode 2 is approximately half of that for Mode 1. 
Previous theoretical work has demonstrated that the frequencies of torsional oscillation eigenmodes are not integer multiples in the presence of dissipation\cite{GJC17}~\hspace{-1.5mm}.
Interestingly the spatiotemporal structures seen in \editor{Figs.~\ref{fig:brightness_dmd_s-t}a-d} resemble those illustrated by the eigenmode calculations.
Unveiling the identified wave patterns could provide us with information about the diffusive transition layer. 
}}
  
\end{description}

\vspace{5mm}
\section*{Data availability}

All data are publicly available: the Jupiter magnetic field model \rev{JRM33\cite{Cetal21}}~\hspace{-1.5mm}, the density distribution model\cite{FBLNBWR12}~\hspace{-1.5mm}, the zonal wind profile OPAL2016\cite{TOLL17}~\hspace{-1.5mm}, and the 5$\mu$m brightness data\cite{AFetal19}~\hspace{-1.5mm}. 
The images of the brightness and other files used to generate the zonal timeseries are available at \url{https://github.com/ArrateAntunano/IRTF_5micron_data}.

\vspace{5mm}
\section*{Code availability}

The codes used in this study are available upon request. The codes related to 
the DMD analysis are publicly available\cite{KBBP16,JSN14}~\hspace{-1.5mm}.

\vspace{5mm}
\section*{Acknowledgements}

\rev{The authors are grateful to J.~Connerney for supplying us the JRM model \editor{and to H.~Cao and K.~Sugiyama for helpful supports}.}
K.H. is supported by the Japan Society for the Promotion of Science
 under Grant-in-Aid for Scientific Research (C) No.~20K04106
 and by Foundation of Kinoshita Memorial Enterprise, Japan.
C.A.J. acknowledges support from the UK's Science and Technology Facilities Council, STFC research grant ST/S00047X/1.
A.A. and L.N.F. were supported by a European Research Council 
(ERC)
Consolidator Grant under the European Union's Horizon 2020 research and innovation program, grant agreement number 723890, at the University of Leicester.
A.A. was also supported by the Spanish project PID2019-109467GB-I00 funded by MCIN/AEI/10.13039/50110001103 and by Grupos Gobierno Vasco IT1742-22.
S.M.T. was supported by an ERC 
Advanced Grant under the European Union's Horizon 2020 research and innovation program (grant agreement
D5S-DLV-786780), at the University of Leeds.

\editor{The authors would like to thank the Isaac Newton Institute for Mathematical Sciences, Cambridge, for support and hospitality during the programme DYT2 where work on this paper was undertaken. This work was supported by EPSRC grant no EP/R014604/1.}

\vspace{5mm}
\section*{Author contributions}

K.H. and C.A.J. designed the research, calculated the wave periods, analysed the brightness time series, and wrote the article. 
A.A. and L.N.F. generated the time series data from the infrared images and provided assistance with the analysis and interpretation. 
S.M.T. provided assistance with the use of the data-driven techniques and the interpretation.
All authors read and commented on the manuscript.

\vspace{5mm}
\section*{\editor{Competing interests}}

\editor{
The authors declare no competing interests.
}

\clearpage
\section*{\editor{Tables}}


\begin{table}[ht]
\caption{Comparison of Jupiter's 5\,$\mu$m observation and torsional wave prediction, in planetocentric coordinates.
 Values in the first two columns are taken from Table 4 in ~\cite{AFetal19} for 5\,$\mu$m-brightness ground observations.
 The third column lists the corresponding, normalised cylindrical radii $s$,
 provided that the radiation senses properties at about Jupiter's 1 bar level, $R_\tx{J}$.
 The Alfv\'{e}n speed, $U_\tx{A}$, listed in the fourth column is calculated from the magnetic field model JRM33\cite{Cetal21} and the H/He model density\cite{FBLNBWR12}~\hspace{-1.5mm}. Since
 waves are heavily damped outside $r_\text{diff}= 0.96R_\tx{J}$ the speed is denoted by $^*$ for $s>r_\text{diff}$. 
 The dimensionless wavenumber, $k R_\tx{J}/2\pi$, in the fifth column is converted from interjet spacings
 in planetocentric latitudes from the OPAL 2016 dataset\cite{TOLL17}~\hspace{-1.5mm}: see Table~\ref{table:jets}.
 The equivalent numbers are also computed from the jet speed profile
 by the Lomb-Scargle periodogram (Supplementary Fig.~3)
 and presented in parentheses. 
 The last column lists the half period of the predicted torsional wave.
 A text file containing the wave periods as a function of latitude is available in Supplementary Table~5.}
\label{table:periods}
\vspace{5mm}
\centering
{\footnotesize
\hspace{-5mm}
\begin{tabular}{rc|cccc}
\hline
Region\editor{(abbreviations)}/latitude & Interval 
  & Normalised & Alfv\'{e}n speed & Wavenumber &  Half period  \\
& (years) & cylindrical  &  $U_\tx{A}$ & $k R_\tx{J}/2\pi$ & $T/2 = \pi/U_\tx{A}k$ \\
&       & radius $s/R_\tx{J}$ & (m/sec)  &  &  (years)  \\
\hline
North Temperate Zone\editor{(NTZ)}/30-33$^\circ$N & 8.5$\pm$0.5 
  & 0.839-0.866 & 0.0269-0.0319 & {13.6} (13.0) &  2.6-3.0 (2.7-3.2)\\ 
North Temperate Belt\editor{(NTB)}/21-23$^\circ$N & {\bf 4.8$\pm$0.6}
  & 0.921-0.934 & 0.0233-0.0234 & 10.2 (13.0) &  {\bf 4.6-4.7} (3.6-3.7)\\
North Equatorial Belt\editor{(NEB)}/14-18$^\circ$N & {\bf 4.4$\pm$0.4}
  & 0.951-0.970 & 0.0161$^*$-0.0223 & 15.5 (13.0) &  {\bf 3.2-4.4$^*$ (3.8-5.3$^*$)} \\
South Equatorial Belt\editor{(SEB)}/12-14$^\circ$S & {\bf 4.5$\pm$0.8} 
  & 0.970-0.978 & 0.0111$^*$-0.0161$^*$  & {13.3 (12.8)} & {\bf 5.2$^*$-7.5$^*$ (5.4$^*$-7.8$^*$)}  \\ 
South Equatorial Belt\editor{(SEB)}/12-15$^\circ$S & 8.2$\pm$0.5
  & 0.966-0.978 & 0.0111$^*$-0.0182$^*$  & {13.3 (12.8)} & 4.6$^*$-7.5$^*$ (4.7$^*$-7.8$^*$)  \\ 
South South Temperate Belt/36-41$^\circ$S&  \rev{~~7$\pm$0.4} 
  & 0.755-0.809 & 0.0411-0.0782 & {11.2} (12.8) & 1.3-2.4 (1.1-2.1) \\ 
\hline
\end{tabular}
}
\editor{
\small Values that match the observation within error bars (the second column) are indicated in bold.
}
\end{table}

\clearpage


\begin{table}[ht]
\caption{Extreme values of the zonal wind profile between planetocentric latitudes $40^\circ$N to $40^\circ$S taken from \editor{the OPAL 2016} dataset\cite{TOLL17}~\hspace{-1.5mm}.
The latitudes of the extrema, the wind speeds attained there, and the corresponding $s/R_\tx{J}$ are shown. The wavenumber obtained by fitting a half wavelength between adjacent extrema is also shown.
The dominant wavenumber peak from a Lomb-Scargle analysis of the dataset between $\pm 40^\circ$ is shown in brackets.}
\label{table:jets}
\vspace{3mm}
\centering
{\small
\begin{tabular}{ccccc}
 \hline
 Band name \editor{(abbreviations)} & Jet latitude & Jet speed & Normalised cylindrical & Wavenumber \\
                                    &              &     (m/sec) & radius  $s/R_\tx{J}$   & $k R_\tx{J}/2\pi$ \\
 \hline
       & 38.8$^\circ$N &  ~~22.7  &  0.779 &  \\
 \raisebox{1ex}[0ex][1ex]{North North Temperate Zone \editor{(NNTZ)}}
       & 35.5$^\circ$N & $-$20.4  &  0.814 & \raisebox{1ex}[0ex][1ex]{14.1}  \\
 \raisebox{1ex}[0ex][1ex]{North North Temperate Belt \editor{(NNTB)}}
       & 31.7$^\circ$N &  ~~34.4  &  0.851 & \raisebox{1ex}[0ex][1ex]{13.6} \\
 \raisebox{1ex}[0ex][1ex]{North Temperate Zone \editor{(NTZ)}} 
       & 27.9$^\circ$N & $-$33.2  &  0.884 & \raisebox{1ex}[0ex][1ex]{15.1} \\
 \raisebox{1ex}[0ex][1ex]{North Temperate Belt \editor{(NTB)}} 
       & 21.1$^\circ$N &   158~~  &  0.933 & \raisebox{1ex}[0ex][1ex]{10.2} \\
 \raisebox{1ex}[0ex][1ex]{North Tropical Zone \editor{(NTrZ)}}
       & 15.1$^\circ$N & $-$19.5  &  0.965 & \raisebox{1ex}[0ex][1ex]{15.5} \\
 \raisebox{1ex}[0ex][1ex]{North Equatorial Belt \editor{(NEB)}} 
       & ~6.7$^\circ$N &  ~~97.4  &  0.993 & \raisebox{1ex}[0ex][1ex]{18.1} \\
                               & & & & (13.0) \\
 \hline
                               & ~6.3$^\circ$S &   147~~  &  0.994 & \\ 
 \raisebox{1ex}[0ex][1ex]{South Equatorial Belt \editor{(SEB)}} 
       & 17.0$^\circ$S & $-$54.6  &  0.956 & \raisebox{1ex}[0ex][1ex]{13.3} \\ 
 \raisebox{1ex}[0ex][1ex]{South Tropical Zone \editor{(STrZ)}}
       & 23.6$^\circ$S &  ~~41.9  &  0.916 & \raisebox{1ex}[0ex][1ex]{12.4} \\
 \raisebox{1ex}[0ex][1ex]{South Temperate Belt \editor{(STB)}} 
       & 29.0$^\circ$S & $-$16.0  &  0.875 & \raisebox{1ex}[0ex][1ex]{12.1} \\
 \raisebox{1ex}[0ex][1ex]{South Temperate Zone \editor{(STZ)}} 
       & 32.4$^\circ$S &  ~~46.7  &  0.844 & \raisebox{1ex}[0ex][1ex]{16.3} \\ 
 \raisebox{1ex}[0ex][1ex]{South South Temperate Belt \editor{(SSTB)}}
       & 35.5$^\circ$S & ~$-$6.5  &  0.814 & \raisebox{1ex}[0ex][1ex]{16.8} \\ 
 \raisebox{1ex}[0ex][1ex]{South South Temperate Zone \editor{(SSTZ)}}
       & 39.7$^\circ$S &  ~~37.0  &  0.769 & \raisebox{1ex}[0ex][1ex]{11.2} \\
                                & & & & (12.8) \\
 \hline
\end{tabular}
}
\end{table}

\clearpage 
\section*{\editor{Figure legends/captions}}


\vspace{15mm}

\vspace{15mm}

\clearpage

\clearpage

\nolinenumbers
\section*{Supplementary information}
\renewcommand{\headrulewidth}{0pt}
\renewcommand{\figurename}{Supplementary Fig. }
\renewcommand{\tablename}{Supplementary Table. }
\setcounter{figure}{0}
\setcounter{table}{0}
\setcounter{page}{1}


\vspace{15mm}
\begin{figure}[ht]
 \begin{tabular}{ll}
   \includegraphics[width=70mm]{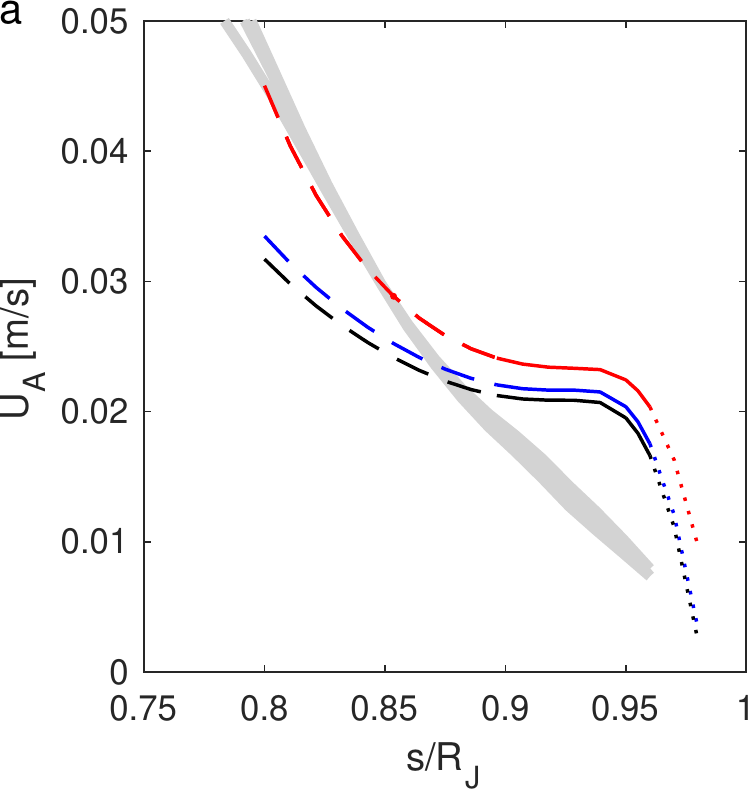} &
   \hspace{10mm} 
   \includegraphics[width=80mm]{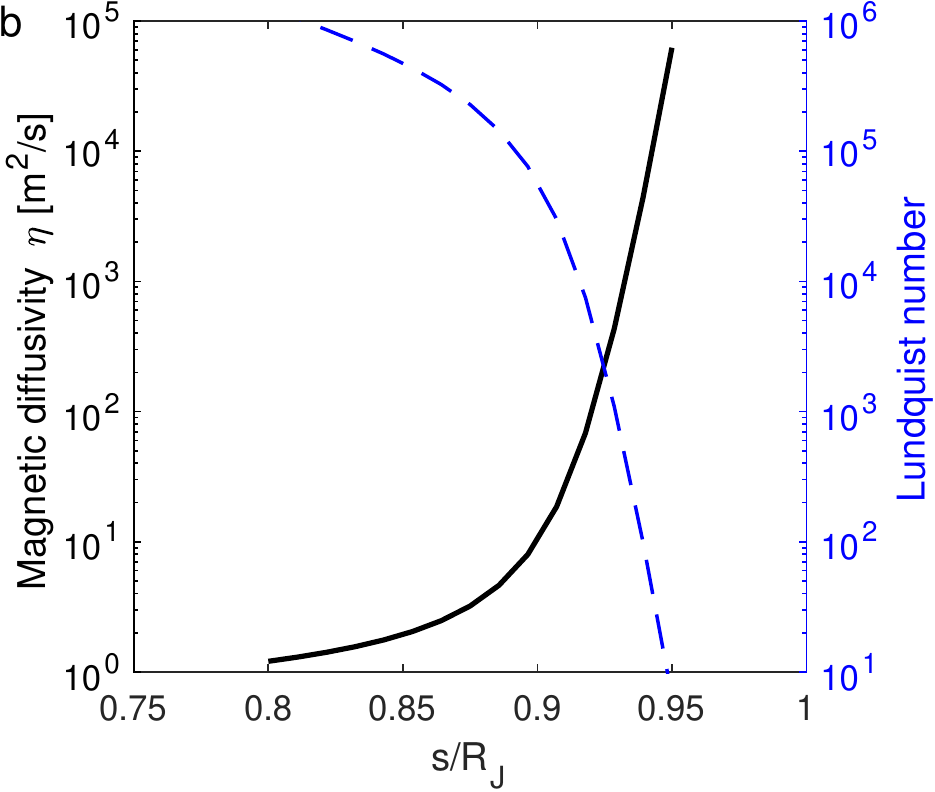}
 \end{tabular}  
   \caption{(a) Profiles of the torsional wave speed $U_\tx{A}$ as a function of the normalised cylindrical radius $s/R_\tx{J}$.
   Red, blue, and black curves represent profiles calculated with the magnetic model JRM33 through degree $n= 18$,
    JRM33 through $n= 10$, and JRM09 through $n= 10$, respectively. 
   Inside $s/R_\tx{J} = 0.9$ (dashed curves) the waves can exist but the speed predictions based on the potential field assumption could overestimate: see details in the main text.
   Outside $s/R_\tx{J} = 0.96$ (dotted curves) the magnetic waves are heavily damped. 
   The case with JRM33 through $n= 18$ is adopted in Table~\ref{table:periods}, all Figures, and Supplementary Table~\ref{table:periods-lat}. 
   Also, grey curves display $U_\tx{A}$ profiles obtained in Jovian dynamo simulations~\cite{TJT15}~\hspace{-1.5mm},
   where the nondimensional speeds are scaled so that they would match the JRM33 speed at $s=0.85 R_\tx{J}$.
   (b) Profiles of the magnetic diffusivity $\eta$ (black solid curve in left ordinate),
   \editor{adapted from }\cite{FBLNBWR12}~\hspace{-1.5mm},
   and of the Lundquist number (blue dashed in right ordinate)
   representing the ratio of the Alfv\'{e}n wave speed to the diffusive one $\eta/D$
   with $D$ being a characteristic length of the conducting region. 
   Here we evaluate the speeds at $z=0$, and suppose \editor{$D=0.2 R_\tx{J}$} as a length scale.} 
 \label{fig:UA_profile}
\end{figure}


\clearpage
\begin{figure}[ht]
\hspace{-10mm}
 \begin{tabular}{rll}
  a \hspace{17mm}  b \hspace{17mm}  c \hspace{17mm}  &   d  &   e  \\
  \includegraphics[height=44mm]{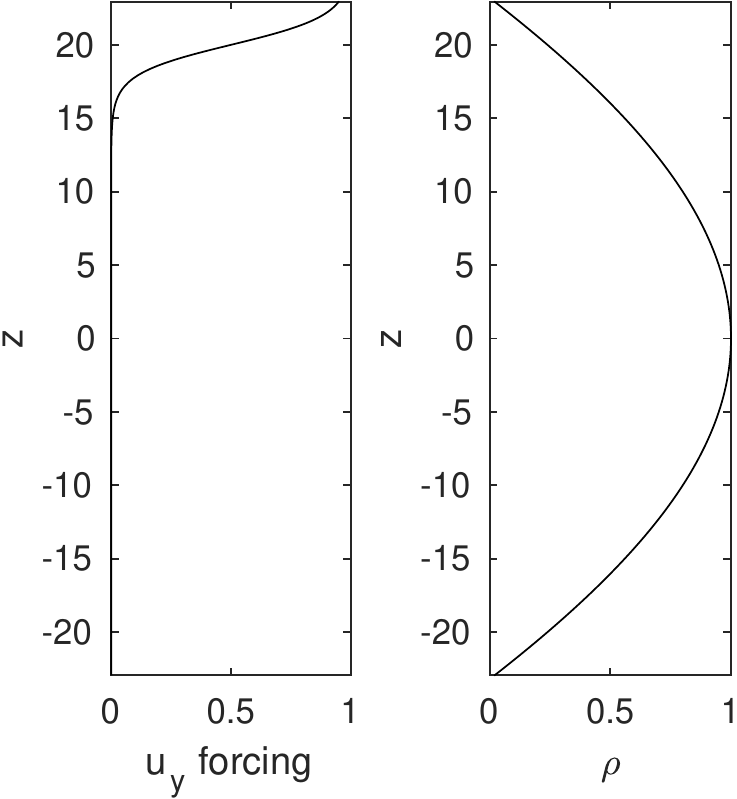}     
  \includegraphics[viewport=0mm 0mm 38mm 81mm,clip,height=44mm]{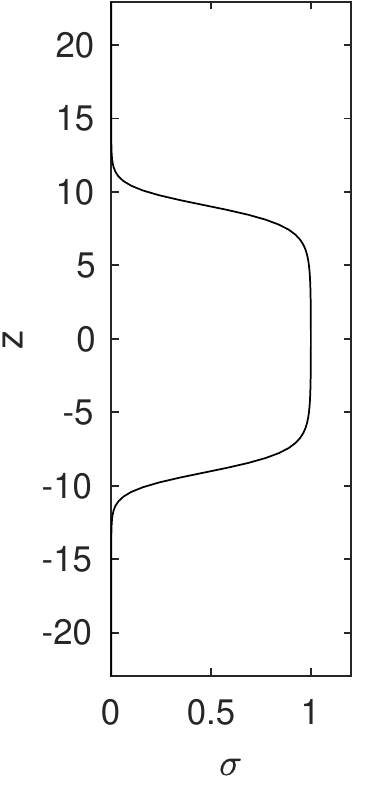} &  
  \includegraphics[viewport=0mm 0mm 110mm 82mm,clip,height=44mm]{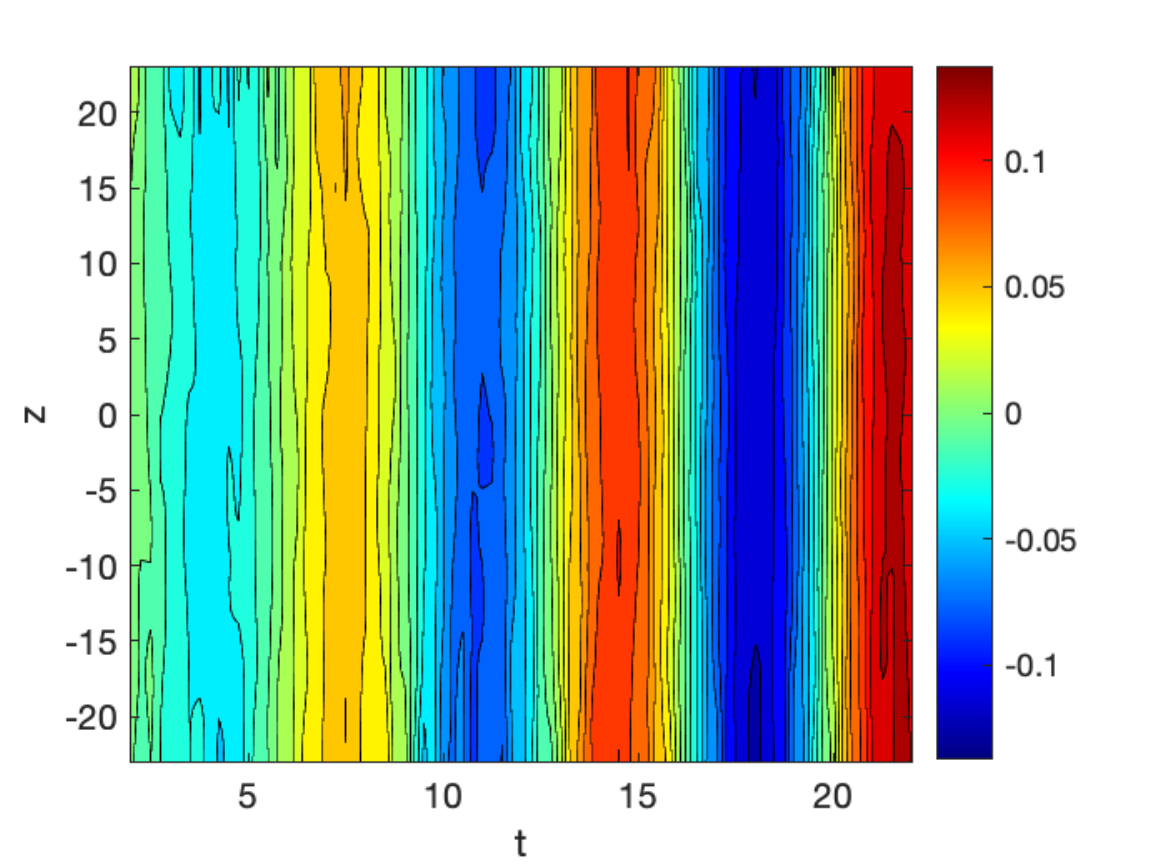} &     
  \includegraphics[viewport=0mm 0mm 110mm 82mm,clip,height=44mm]{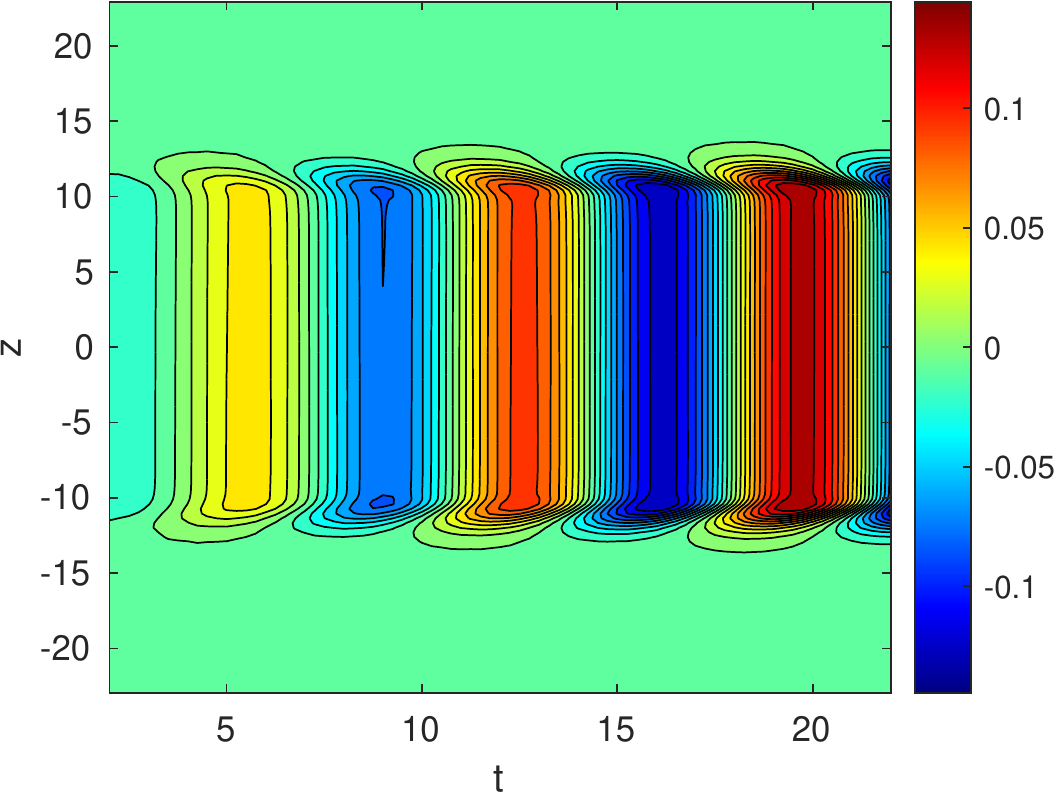} \\    
  \vspace{5mm}
  \\
  f \hspace{17mm} &  g  &  h \\
  \includegraphics[viewport=38mm 0mm 76mm 81mm,clip,height=44mm]{./figureS2c-f} &   
  \includegraphics[viewport=0mm 0mm 110mm 82mm,clip,height=44mm]{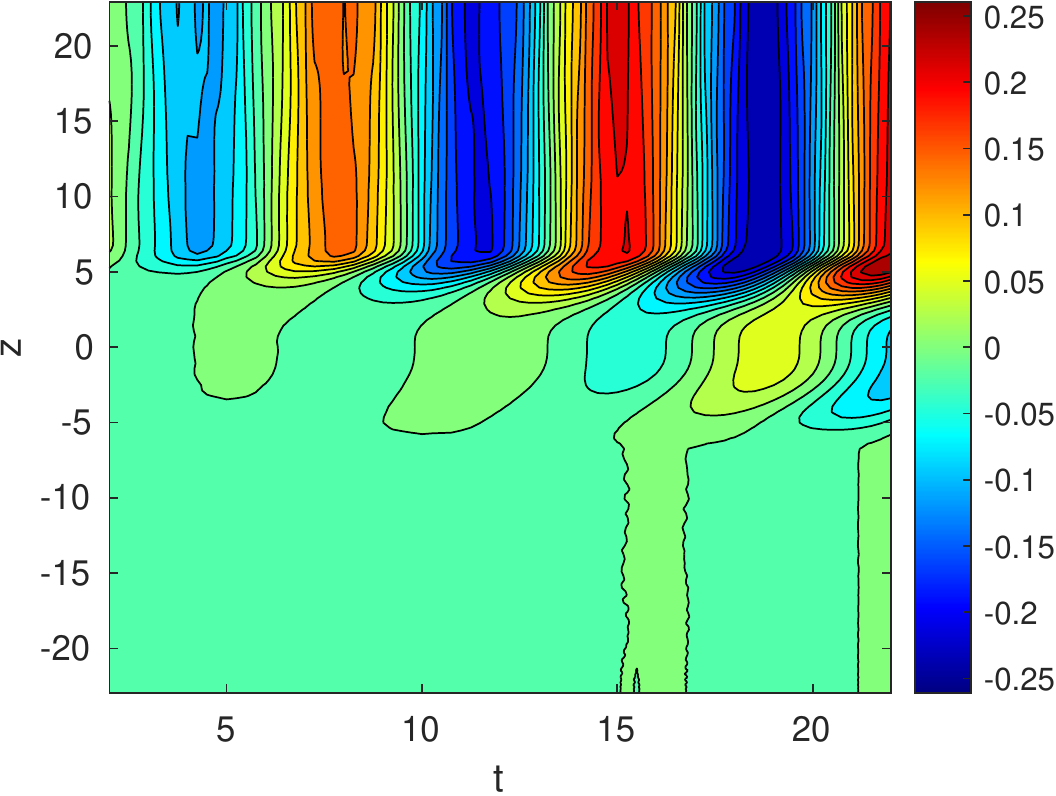} &     
  \includegraphics[viewport=0mm 0mm 110mm 82mm,clip,height=44mm]{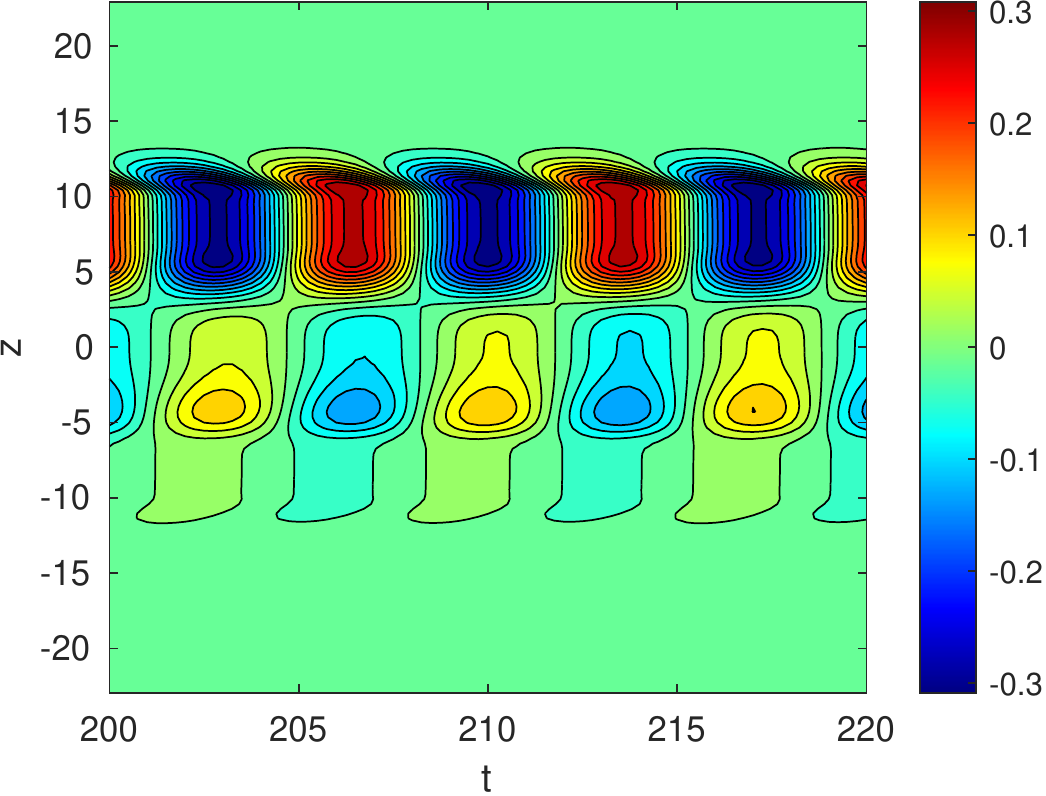}       
 \end{tabular}
 \caption{Forward simulations of torsional oscillations driven by a zonal wind variation. For radial wavenumber of 1. \editor{The unit of length is 1,091\,km and the annulus $z$-coordinate runs from $-$25,031\,km to $+$25,031\,km, which is $-$22.94 to $+$22.94 in dimensionless units.
 (a) The $z$ profile of the assumed wind forcing, $F(z) = 0.5 + 0.5 \tanh{( (z - z_\textrm{f})/h_\textrm{f} )}$ where $z_\textrm{f} = 20$ and $h_\textrm{f} = 2$. (b) Background density $\rho$ profile. (c) Electrial conductivity $\sigma$ profile, $0.5 + 0.5 \tanh{( (z_\textrm{cond}^2 - z^2)/h_\textrm{cond}^2 )}$ where $z_\textrm{cond} = 9$ and $h_\textrm{cond}=5$.} 
 (d) Evolution of the zonal velocity and (e) of the magnetic fluctuation in the absence of stably stratified layers. The wind variation given in the northern surface immediately spreads throughout $z$ to excite equatorially-symmetric magnetic oscillations. 
 \editor{(f) The $z$ profile of the background $g_z dS_0/dz$ to model a stably stratified layer, $(z/z_\textrm{buoy})^2 (0.5 + 0.5 \tanh{( (z_\textrm{buoy}^2 - z^2)/h_\textrm{buoy}^2)}$ where $z_\textrm{buoy}=6$ and $h_\textrm{buoy}=2$. The point in the annulus where $z=z_\textrm{buoy}=6$ lies at a depth of 4,300\,km below the surface}. 
 (g-h) Similarly to figures d-e but in the presence of the stratified layer. In figure h the magnetic fluctuation in the final stage is shown. 
 The presence of the stratified layer inhibits the forcing from penetrating and gives rise to weaker signatures near the equator and in the southern hemisphere.}
 \label{fig:forward_TO}
\end{figure}


\clearpage
\vspace{15mm}
\begin{figure}[ht]
 \centering
   \includegraphics[width=135mm]{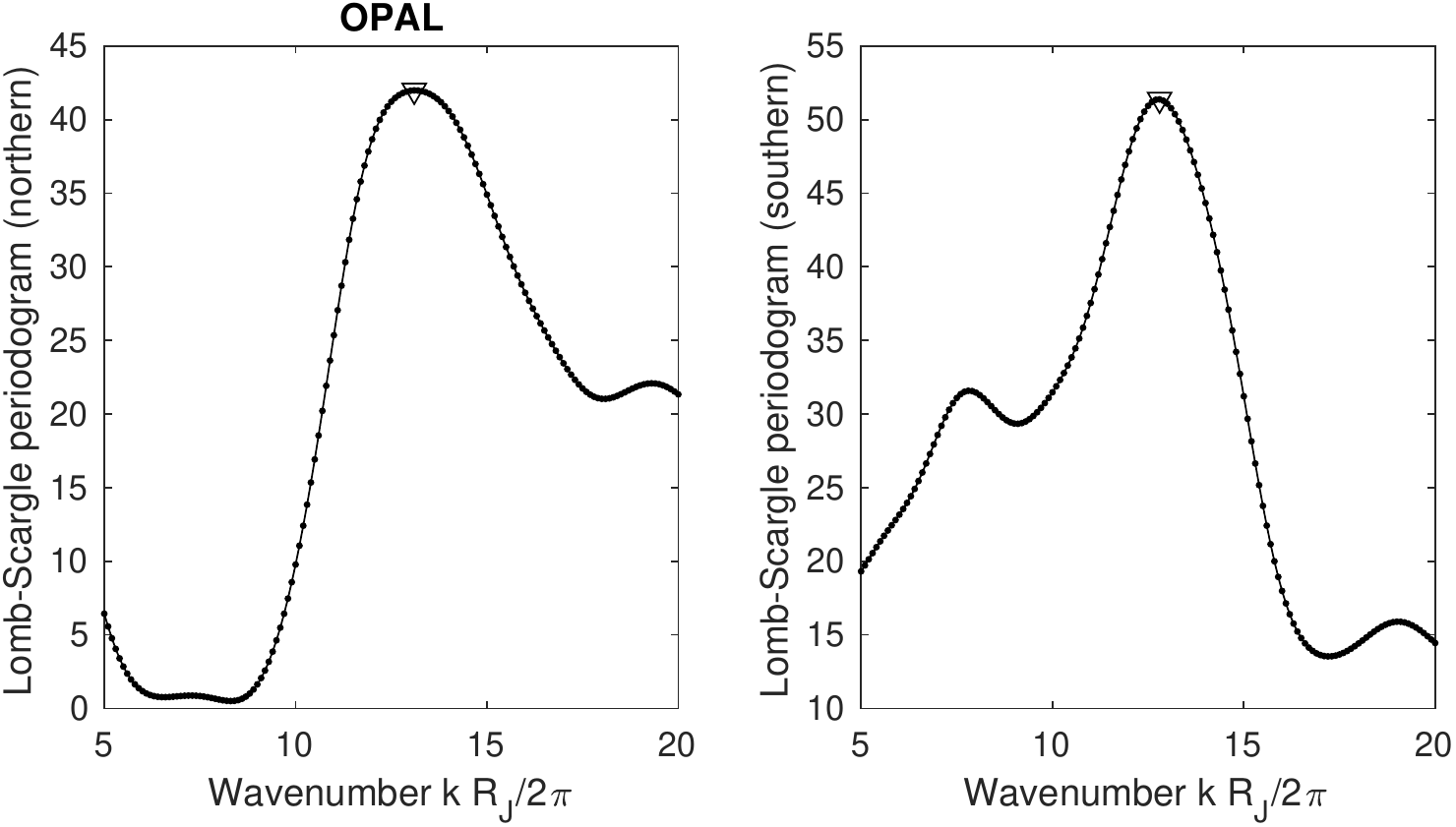} 
 \caption{Lomb-Scargle periodogram of the northern (left) and southern (right) jet profiles as a function of $s/R_\tx{J}$,
  based on the OPAL 2016 data set of Tollefson {\it et al.}~\cite{TOLL17}~\hspace{-1.5mm}.
 Only the data between latitudes $\pm 40^\circ$ are used.  
 The very dominant peaks, marked with triangles, correspond to wavenumbers $k R_\tx{J}/2 \pi = 13.0$ for the northern jets
 and 12.8 for the southern jets. As expected, very similar peaks were obtained from the perijove3 data set of \editor{\cite{TOLL17}}~\hspace{-1.5mm}. } 
\label{fig:jets_spectra}
\end{figure}


\clearpage

\begin{figure}[ht]
 \includegraphics[width=165mm]{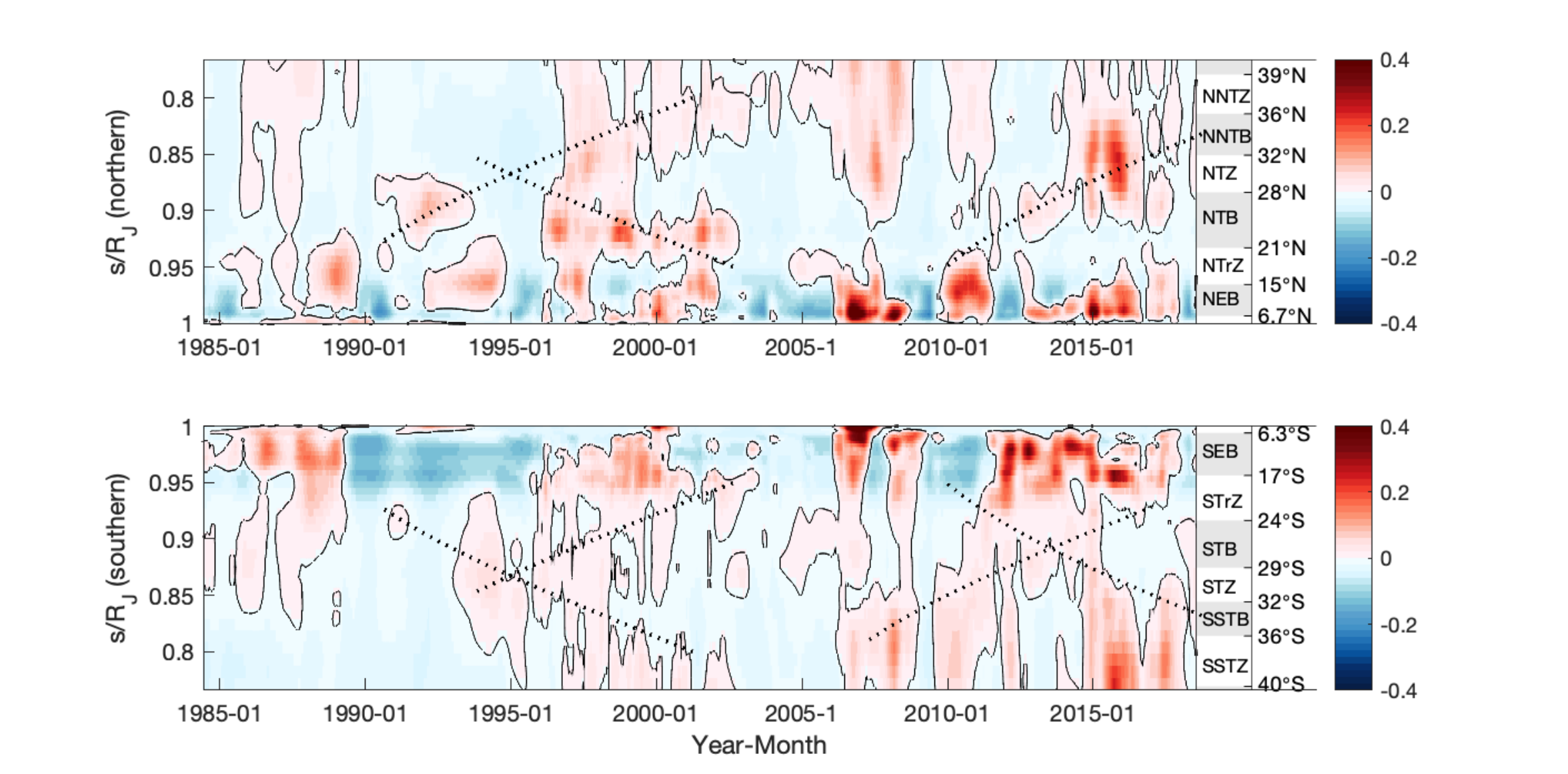}
\caption{Anomaly $L'(\theta,t)$ of Jupiter's 5$\mu$m brightness (\editor{adapted} from Antu\~{n}ano {\it et al.}~\cite{AFetal19}),
 which is interpolated with cubic polynomials,
 smoothed by running averages with a window size of 100 days. 
The observed 5$\mu$m radiances were all scaled to 46-48$^\circ$S ($0.695 \le s/R_\tx{J} \le 0.669$), a relatively quiescent region identified by ~\cite{AFetal19}~\hspace{-1.5mm}.
 The latitudinal dataset is represented with respect to the cylindrical radius, $s/R_\tx{J}$,
 in northern (top panel) and southern (bottom) hemispheres.
 The jet latitudes and bands (Table~\ref{table:jets}) are presented on the right side.
 The dotted curves indicate arbitrary \editor{phase} paths of 
 torsional waves, with wave speed given by $U_\tx{A}$, similarly to \editor{Fig.~\ref{fig:brightness_dmd_s-t}} (details in Methods).}
 \label{fig:brightness}
\end{figure}


\vspace{15mm}
\begin{figure}[hb]
\hspace{15mm}
 \begin{tabular}{rl}
  \includegraphics[width=55mm]{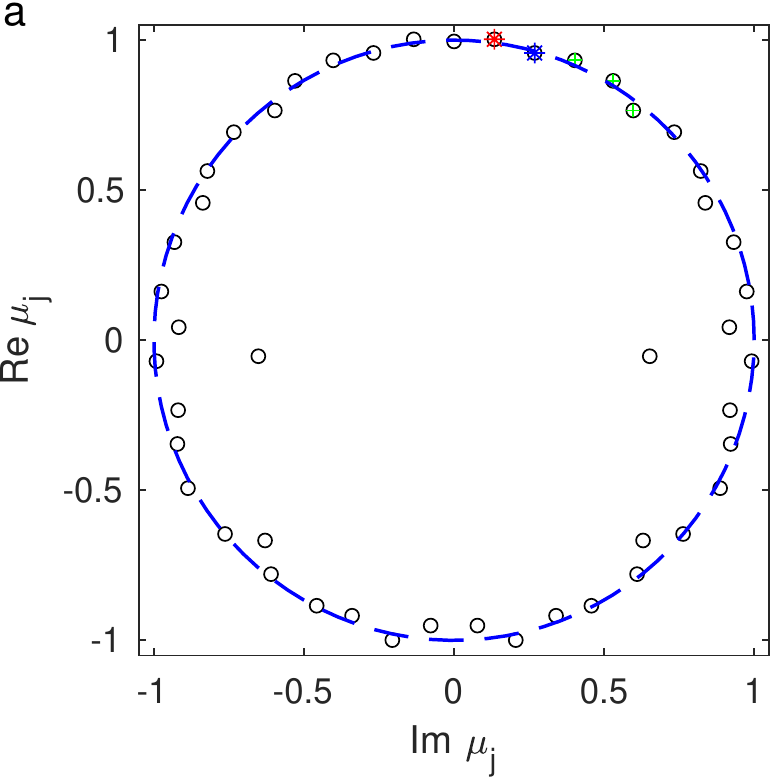} &
  \hspace{5mm}
  \includegraphics[width=55mm]{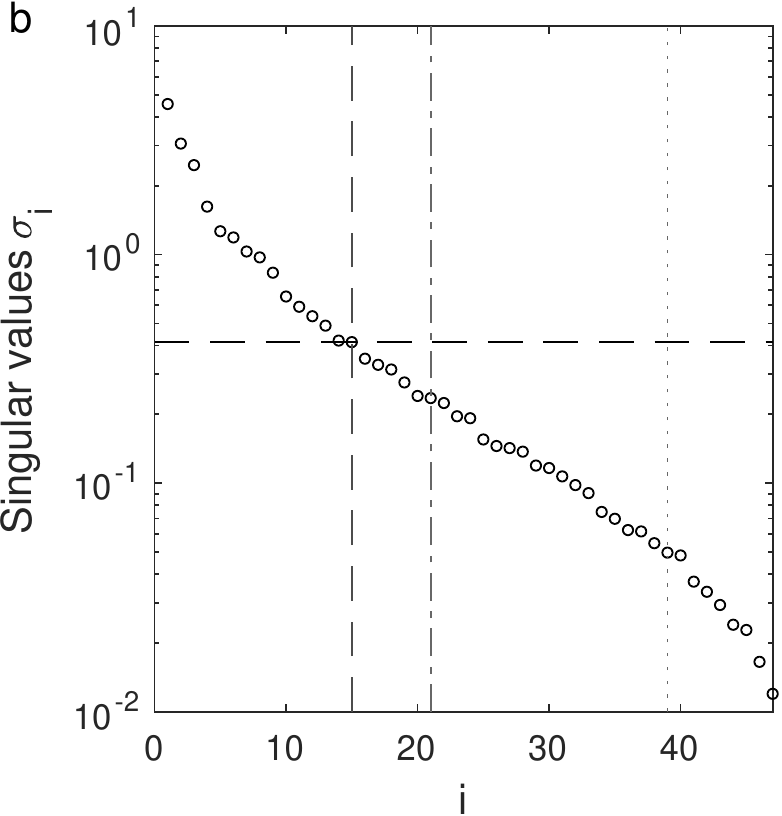} \\
  \\
  \includegraphics[height=80mm]{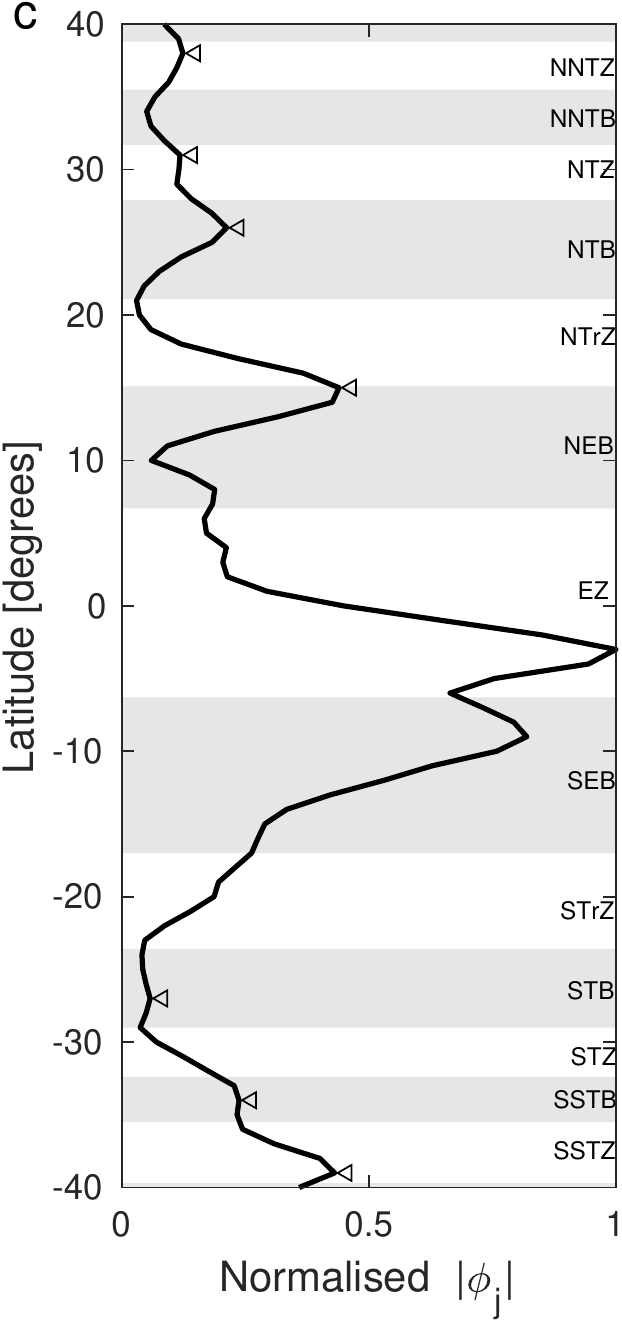}  \hspace{5mm} &
  \hspace{5mm}
  \includegraphics[height=85mm]{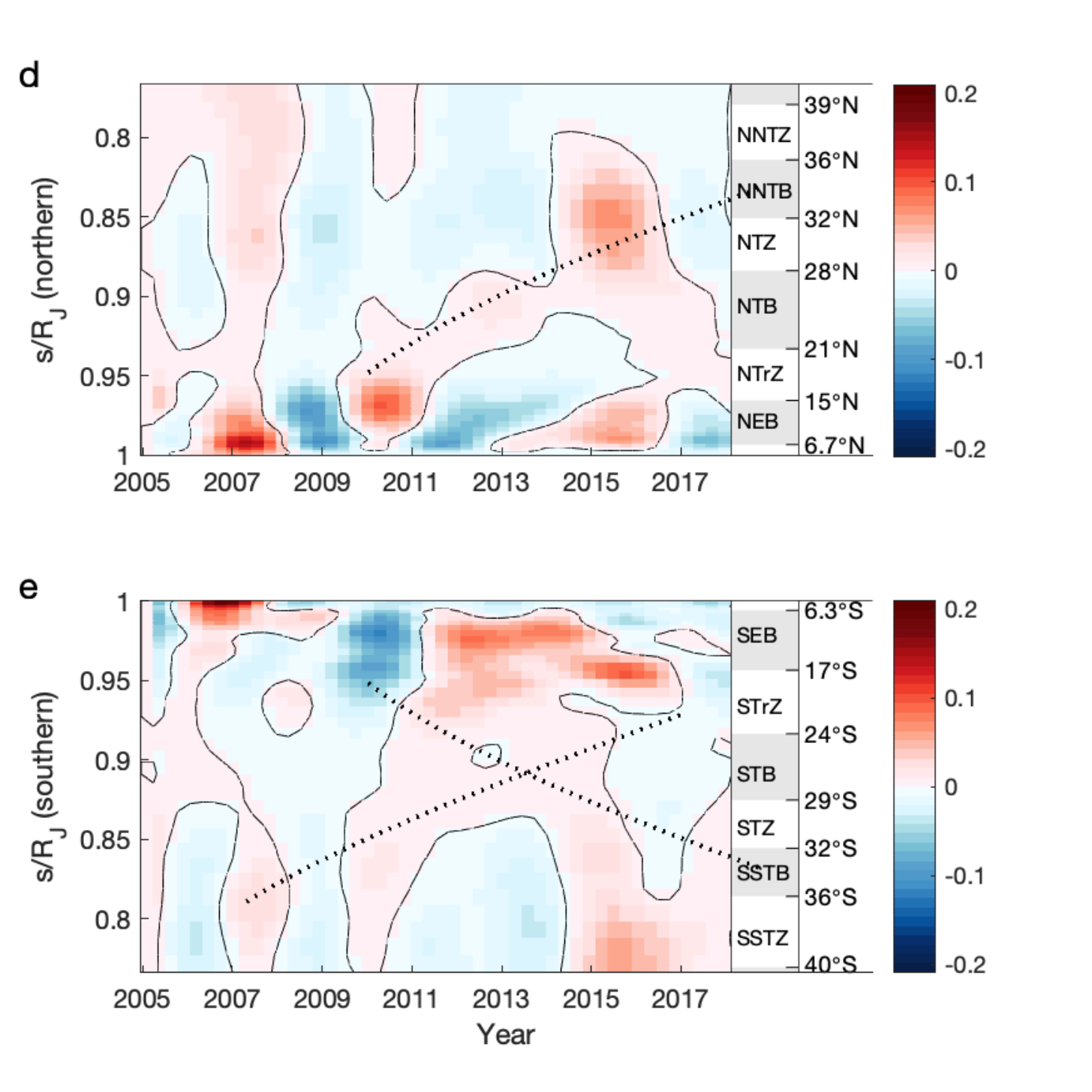} 
 \end{tabular}
 \caption{Additional plots to the DMD of $L' (\theta,t)$ as shown in \editor{Figs.~\ref{fig:brightness_dmd}-\ref{fig:brightness_dmd_s-t}}. 
 (a) DMD eigenvalues $\mu_j$ presented over a unit circle in complex space. 
  Modes 1 and 2 are highlighted in red and blue, respectively;
  three more modes in the wave frequency range
    (the shaded region of Fig.~\ref{fig:brightness_dmd}a-b) are depicted in green.
 (b) Singular values $\sigma_j$ against $j$.
 The vertical and horizontal dashed lines indicate the optimal hard threshold for unknown random noise \cite{GD14}~\hspace{-1.5mm}.
 Thresholds to retain 90\% and 99 \% energy of the total variance are represented by the vertical dashed-dotted and dotted lines, respectively.
 (c) Normalised, latitudinal profiles of the superposition of Mode 1, Mode 2, and the three modes,
   as in Fig.~\ref{fig:brightness_dmd}c.
 (d-e) Reconstructed spatiotemporal structures of the superposed, five modes 
  in the northern (d) and southern hemispheres (e).
 Dotted curves indicate the \editor{phase} paths \editor{that are presented in Fig.~\ref{fig:brightness_dmd_s-t}. 
 In (a) and (c-e) 
 the full 47 modes are retained, as in \editor{Figs.~\ref{fig:brightness_dmd}-\ref{fig:brightness_dmd_s-t}}.} }
 \label{fig:brightness_dmd_sup}

\end{figure}

\clearpage

 
\begin{table}[ht]
 \begin{center}
 \caption{Convective velocity estimates as a function of depth below Jupiter's surface, with the thermodynamic quantities\cite{FBLNBWR12} required for formulae (\ref{eq:showman}) and (\ref{eq:cia}). }
 \label{table:convective_vel}
\vspace{3mm}
\def~{\hphantom{0}}
\begin{tabular}{ccccccc}
\hline
Depth  & $U_\text{Show}$ & $U_\text{CIA}$  & H & $\rho$   & $\alpha$  & $c_p$ \\
 (km)  & (cm/sec)  & (cm/sec) & (km) & (kg\,m$^{-3}$) &  ($^\circ$K$^{-1}$) & (J\,kg$^{-1}$\,$^\circ$K$^{-1}$) \\
\hline
~420 & 7.2 & 20.6  & ~158  & ~11.3 &  $9.5 \times 10^{-4}$  & $1.22 \times 10^4$ \\
~979 & 2.1 & ~9.3  & ~367  & ~49.7 &  $3.9 \times 10^{-4}$  & $1.29 \times 10^4$ \\
1399 & 1.3 & ~6.9  & ~566  & ~84.8 &  $2.6 \times 10^{-4}$  & $1.34 \times 10^4$ \\ 
1958 & 0.8 & ~5.1  & ~805  & 132~  &  $1.7 \times 10^{-4}$  & $1.40 \times 10^4$ \\ 
2578 & 0.6 & ~4.3  & 1094  & 177~  &  $1.4 \times 10^{-4}$  & $1.54 \times 10^4$ \\ 
\hline
\end{tabular}
\end{center}
\end{table}


\clearpage
\vspace{15mm}
\begin{table}[hb]
\caption{Performance of the DMD for different values of latitudinal ranges, temporal duration, coordinate delay $d$, and mode truncation $r$.
In the fourth column the 90\%-energy, 99\%-energy, and optimal hard thresholds are indicated in parentheses.
The fifth column to confirm the absence of zero eigenvalues, $\mu_j = 0$.
 The root mean square error (RMSE) 
 is computed for two formulae of amplitude, $\mib{b}_\text{I}$ and $\mib{b}_\text{P}$. 
 The last four columns list half periods, $T_j/2=\pi/\text{Im}\lambda_j$, and quality factors, $Q_j=|\text{Im}\lambda_j/2\text{Re}\lambda_j|$, of two modes that have lowest nonzero frequencies amongst the obtained DMD modes. 
 The half periods are presented in years. 
 \editor{Cases} corresponding to \editor{Figs.~\ref{fig:brightness_dmd}-\ref{fig:brightness_dmd_s-t}} and Supplementary Fig.~\ref{fig:brightness_dmd_sup} are indicated in bold.} 
\label{table:DMD_errors}
\vspace{3mm}
\hspace{-4mm}
{\footnotesize
\begin{tabular}{llrl l ll rlrl}
\hline
 Latitudes & Duration & 
 $d$ & 
 ~~$r$ &  zero $\mu_j$ &
  & \hspace{-8mm}RMSE &  
  & \hspace{-8mm}Mode 1 &  & \hspace{-8mm}Mode 2 \\
  &  & 
  &  &   &
 for $\mib{b}_\text{I}$ & for $\mib{b}_\text{P}$ & 
 $T_j/2$ & ~~$Q_j$  &  $T_j/2$ & ~~$Q_j$ \\
  &  & 
  &  &   &
  &  &  
 (\editor{years}) &  &  (\editor{years}) &  \\
 \hline
 {\bf 40$^\circ$S-40$^\circ$N} & {\bf Mar 2005-Jul 2018} &  
   {\bf 1}& {\bf 47} ($= r_\text{f}$) & {\bf no} & 6.47e-04 & {\bf 1.14e-04}  & {\bf ~6.46}& {\bf ~~5.80~} &  {\bf 3.15}& {\bf 22.9~} \\
  &&    2  & 46 ($= r_\text{f}$) &yes & 4.34e-04 & 1.18e-04  &      ~6.73 & ~14.6~~ &      3.06 & 26.3~ \\
 &&     1 & 39 (99\%)      & no & 7.24e-04 & 7.69e-04  &      ~6.56 & ~~4.91~ &       3.20 & ~7.77 \\
 &&     1 & 21 (90\%)      & no & 1.14e-03 & 1.17e-03  &      22.1~ & ~~0.204 &       3.13 & ~1.70 \\
 &&     1 & 15 (hard)      &yes & 1.23e-03 & 1.19e-03  &      ~3.32 & ~~1.23~ &       2.06 & ~1.79 \\
 &&     1 & ~5             & no & 1.23e-03 & 1.23e-03  &      ~5.40 & ~~0.443 &       1.64 & ~5.59  \\
 &&     1 & ~3             & no & 1.31e-03 & 1.31e-03  &      ~6.57 & ~~0.497 & --- \\
\\
 40$^\circ$S-40$^\circ$N & Nov 2003-Jul 2018 &  
        1 &  52 ($= r_\text{f}$) &  no &  1.80e-03 & 7.08e-05 &     ~6.24 & ~~5.95~ &      3.09 & 39.4~ \\
 70$^\circ$S-70$^\circ$N & Mar 2005-Jul 2018 & 
        1 &  47 ($= r_\text{f}$) & no & 2.29e-04 & 7.02e-05  &      ~6.45 & ~14.4~~ &       3.12 & 12.7~ \\
 16-40$^\circ$N/S       & Mar 2005-Jul 2018 & 
        1 &  47 ($= r_\text{f}$) &  no &  9.03e-04 & 5.27e-05 &     ~6.32 & ~~6.99~ &  3.20 & 26.8~ \\
%
\hline
\end{tabular}
}
\end{table}


\begin{table}[ht]
  \caption{Comparison of length scales in surface zonal winds (Table~\ref{table:jets}) and in filtered brightness (Fig.~\ref{fig:brightness_dmd}c and Supplementary Fig.~\ref{fig:brightness_dmd_sup}c).
   In the first column squares mark latitudinal bands in which the 5-$\mu$m brightness variation \cite{AFetal19} is consistent with the torsional wave prediction (Table~\ref{table:periods}):
   daggers indicate zonal bands when the surface wind noticeably varies \cite{TOLL17}~\hspace{-1.5mm}. 
   Columns 5-10 list latitudes at which local peaks and minima of the DMD eigenfunctions $|\phi_j(\theta)|$ are identified by the \editor{Matlab function findpeaks}.
   The jet latitudes are indicated in bold when the superposition of Modes 1-2 endorses corresponding peaks.
   Only latitudes $\ge 14^\circ$ are listed.}
  \label{table:jets_vs_dmd}
\vspace{3mm}
\hspace{-3mm}
{\footnotesize
\begin{tabular}{lccc|rl|rl| r|r}
 \hline
 Band  & Jet latitude & Jet speed & Radius         
       &                 & \hspace{-8mm}Mode 1 &              & \hspace{-8mm}Mode 2 &  Modes 1-2  &  Modes 1-5 \\
       &  ($^\circ$)   &     (m/sec) & $s/R_\tx{J}$   
       &  peaks ($^\circ$)  & minima ($^\circ$) & peaks ($^\circ$) & minima ($^\circ$) &  peaks ($^\circ$) & peaks ($^\circ$) \\ 
 \hline
                                & {\bf 38.8N} &  ~~22.7 &0.779 
       &  -- &      &  38N &                                & {\bf 39N}&   38N \\
 \raisebox{1ex}[0ex][1ex]{NNTZ} &      35.5N  & $-$20.4  &  0.814
       &  -- &  35N &   -- &  35N &       -- &    -- \\
 \raisebox{1ex}[0ex][1ex]{NNTB} & {\bf 31.7N} &  ~~34.4 &0.851 
       & 31N &      &  31N &      & {\bf 31N}&   31N \\
 \raisebox{1ex}[0ex][1ex]{NTZ}  & {\bf 27.9N} & $-$33.2 &0.884 
       &  -- &      &   -- &  26N & {\bf 28N} &  26N \\
 \raisebox{1ex}[0ex][1ex]{NTB$^{\square \dagger}$}  & 21.1N &   158~~  &  0.933 
       &  -- &  20N &  23N &  \raisebox{-1ex}[0ex][-1ex]{21N} &      -- &    -- \\
 \raisebox{1ex}[0ex][1ex]{NTrZ} & {\bf 15.1N} & $-$19.5 &0.965 
       &  -- &      &  14N &      &{\bf14N}   & 15N \\
 \raisebox{1ex}[0ex][1ex]{NEB$^\square$}  \\ 
 \hline
  & \\
 \raisebox{1ex}[0ex][1ex]{SEB$^{\square \dagger}$}  & {\bf 17.0S} & $-$54.6 &0.956 
       &  17S &                              &   -- &                               &  {\bf 18S}&   -- \\ 
 \raisebox{1ex}[0ex][1ex]{STrZ} &      23.6S &  ~~41.9  &  0.916 
       &   -- &  27S &   -- &  25S &   --  &   -- \\
 \raisebox{1ex}[0ex][1ex]{STB}  & {\bf 29.0S} & $-$16.0 &0.875 
       &  29S &      &  28S &      & {\bf 28S}&  27S \\
 \raisebox{1ex}[0ex][1ex]{STZ}  & {\bf 32.4S} &  ~~46.7 &0.844 
       &   -- &  32S &      &  31S & {\bf 33S}&   \\ 
 \raisebox{1ex}[0ex][1ex]{SSTB} &      35.5S & ~$-$6.5  &  0.814 
       &   -- &      & \raisebox{1ex}[0ex][1ex]{34S} & 36S  &  &  \raisebox{1ex}[0ex][1ex]{34S} \\ 
 \raisebox{1ex}[0ex][1ex]{SSTZ} &      39.7S &  ~~37.0  &  0.769 
       &   -- &      &  39S &   &   --   &  39S \\
 \hline
\end{tabular}
}
\end{table}


\clearpage
\begin{table}[ht]
  \caption{Comparison of torsional wave predictions between different magnetic field models, JRM09 and JRM33. Here the model coefficients in spherical harmonics through degree $n= 10$ are taken. 
  (cf. JRM33 through $n = 18$ in Table~\ref{table:periods}.)}
 \label{table:periods_deg10}
\vspace{3mm}
\hspace{-5mm}
{\footnotesize
\begin{tabular}{rcc|cc|cc}
\hline
Region/latitude 
  & Normalised & Wavenumber & JRM33: Alfv\'{e}n  & : Half period & JRM09: Alfv\'{e}n & : Half period  \\
 & cylindrical & $k R_\tx{J}/2\pi$  & speed $U_\tx{A}$ & $T/2 = \pi/U_\tx{A}k$   &  speed $U_\tx{A}$ & $T/2 = \pi/U_\tx{A}k$ \\
 & radius $s/R_\tx{J}$ &             & (m/sec)  &  (\editor{years})  & (m/sec)  &  (\editor{years})  \\
\hline
\hline
NTZ/30-33$^\circ$N 
  & 0.866-0.839 & 13.6 (13.0)  &  0.024-0.027 & 3.0-3.4 (3.2-3.6)  &  0.023-0.026 &  3.1-3.5 (3.3-3.7) \\
                         
NTB/21-23$^\circ$N 
  & 0.934-0.921 & 10.2 (13.0)  &  0.022       & {5.0} (3.9)    &  0.021       &  {5.2} (4.1)\\
NEB/14-18$^\circ$N 
  & 0.970-0.951 & 15.5 (13.0)  &  0.012$^*$-0.020 & {3.7-6.5$^*$(4.2-7.2$^*$)}  & 0.011$^*$-0.019 & {3.7-6.5$^*$(4.4-7.8$^*$)} \\
                          
SEB/12-15$^\circ$S 
  & 0.978-0.966 & 13.3 (12.8)  &  0.0052$^*$-0.015$^*$ & {5.7$^*$-16$^*$(5.9$^*$-17$^*$)}  &  0.0043$^*$-0.014$^*$  & {6.1$^*$-20$^*$(6.3$^*$-20$^*$)} \\ 
SSTB/36-41$^\circ$S 
  & 0.809-0.755 & 11.2 (12.8)  &  0.032-0.046 & 2.2-3.1 (1.9-2.7)  &  0.030-0.043 & 2.3-3.3 (2.0-2.9) \\ 
\hline
\end{tabular}
}
\end{table}


\clearpage
\begin{table}[ht]
  \caption{Listing of the expected torsional wave periods as a function of latitude (cf. Table~\ref{table:periods}). 
  Column 1 lists the latitude. The corresponding normalised cylindrical radius, $s/R_\tx{J}$, and the Alfv\'{e}n wave speed, $U_\tx{A}$, are listed in columns 2 and 3, respectively. Columns 4 and 5 list, respectively, the dimensionless wavenumber, $k R_\tx{J}/2\pi$, converted from interjet spacings and the half period, $T/2 = \pi/U_\tx{A}k$, of the predicted wave. 
  The last two columns list the equivalent numbers but for the case when the wavenumber is computed from the jet speed profile by the Lomb-Scargle periodogram.}
 \label{table:periods-lat}
\vspace{3mm}
\centering
{\footnotesize
\begin{tabular}{rcc cc cc}
\hline
  latitude & $s/R_\text{J}$  & $U_\text{A}$  &  $k R_\text{J}/2\pi$ & $T/2$ & $k R_\text{J}/2\pi$ & $T/2$ \\
  $(^\circ)$  &  & (m/sec) & & (\editor{years}) & & (\editor{years}) \\
\hline
  35 &  0.8192 &  0.03735  &   13.6  &  2.18  &  13.0  &  2.28 \\
  34 &  0.8290 &  0.03433  &   13.6  &  2.37  &  13.0  &  2.48 \\
  33 &  0.8387 &  0.03188  &   13.6  &  2.55  &  13.0  &  2.67 \\
  32 &  0.8480 &  0.02988  &   13.6  &  2.73  &  13.0  &  2.85 \\
  31 &  0.8572 &  0.02825  &   15.1  &  2.60  &  13.0  &  3.02 \\
  30 &  0.8660 &  0.02693  &   15.1  &  2.72  &  13.0  &  3.16 \\
  29 &  0.8746 &  0.02588  &   15.1  &  2.83  &  13.0  &  3.29 \\
  28 &  0.8829 &  0.02506  &   15.1  &  2.93  &  13.0  &  3.40 \\
  27 &  0.8910 &  0.02443  &   10.2  &  4.44  &  13.0  &  3.49 \\
  26 &  0.8988 &  0.02398  &   10.2  &  4.53  &  13.0  &  3.55 \\
  25 &  0.9063 &  0.02367  &   10.2  &  4.59  &  13.0  &  3.60 \\
  24 &  0.9135 &  0.02348  &   10.2  &  4.62  &  13.0  &  3.63 \\
  23 &  0.9205 &  0.02337  &   10.2  &  4.64  &  13.0  &  3.64 \\
  22 &  0.9272 &  0.02333  &   10.2  &  4.65  &  13.0  &  3.65 \\
  21 &  0.9336 &  0.02331  &   15.5  &  3.06  &  13.0  &  3.65 \\
  20 &  0.9397 &  0.02321  &   15.5  &  3.08  &  13.0  &  3.67 \\
  19 &  0.9455 &  0.02290  &   15.5  &  3.12  &  13.0  &  3.72 \\
  18 &  0.9511 &  0.02228  &   15.5  &  3.21  &  13.0  &  3.82 \\
  17 &  0.9563 &  0.02131  &   15.5  &  3.35  &  13.0  &  4.00 \\
  16 &  0.9613 &  0.01996  &   15.5  &  3.58  &  13.0  &  4.27 \\
  15 &  0.9659 &  0.01823  &   18.1  &  3.36  &  13.0  &  4.67 \\
  14 &  0.9703 &  0.01614  &   18.1  &  3.79  &  13.0  &  5.28 \\
  13 &  0.9744 &  0.01375  &   18.1  &  4.45  &  13.0  &  6.20 \\
  12 &  0.9781 &  0.01109  &   18.1  &  5.52  &  13.0  &  7.68 \\
\\
 -12 &  0.9781 &  0.01109  &   13.3  &  7.51  &  12.8  &  7.80 \\
 -13 &  0.9744 &  0.01375  &   13.3  &  6.06  &  12.8  &  6.29 \\
 -14 &  0.9703 &  0.01614  &   13.3  &  5.16  &  12.8  &  5.36 \\
 -15 &  0.9659 &  0.01823  &   13.3  &  4.57  &  12.8  &  4.75 \\
 -16 &  0.9613 &  0.01996  &   13.3  &  4.17  &  12.8  &  4.33 \\
 -17 &  0.9563 &  0.02131  &   13.3  &  3.91  &  12.8  &  4.06 \\
 -18 &  0.9511 &  0.02228  &   12.4  &  4.01  &  12.8  &  3.88 \\
 -19 &  0.9455 &  0.02290  &   12.4  &  3.90  &  12.8  &  3.78 \\
 -20 &  0.9397 &  0.02321  &   12.4  &  3.85  &  12.8  &  3.73 \\
 -21 &  0.9336 &  0.02331  &   12.4  &  3.83  &  12.8  &  3.71 \\
 -22 &  0.9272 &  0.02333  &   12.4  &  3.83  &  12.8  &  3.71 \\
 -23 &  0.9205 &  0.02337  &   12.4  &  3.82  &  12.8  &  3.70 \\
 -24 &  0.9135 &  0.02348  &   12.1  &  3.90  &  12.8  &  3.68 \\
 -25 &  0.9063 &  0.02367  &   12.1  &  3.87  &  12.8  &  3.65 \\
 -26 &  0.8988 &  0.02398  &   12.1  &  3.82  &  12.8  &  3.61 \\
 -27 &  0.8910 &  0.02443  &   12.1  &  3.75  &  12.8  &  3.54 \\
 -28 &  0.8829 &  0.02506  &   12.1  &  3.65  &  12.8  &  3.45 \\
 -29 &  0.8746 &  0.02588  &   12.1  &  3.54  &  12.8  &  3.34 \\
 -30 &  0.8660 &  0.02693  &   16.3  &  2.52  &  12.8  &  3.21 \\
 -31 &  0.8572 &  0.02825  &   16.3  &  2.41  &  12.8  &  3.06 \\
 -32 &  0.8480 &  0.02988  &   16.3  &  2.27  &  12.8  &  2.90 \\
 -33 &  0.8387 &  0.03188  &   16.8  &  2.07  &  12.8  &  2.71 \\
 -34 &  0.8290 &  0.03433  &   16.8  &  1.92  &  12.8  &  2.52 \\
 -35 &  0.8192 &  0.03735  &   16.8  &  1.76  &  12.8  &  2.32 \\
\hline
\end{tabular}
}
\end{table}

\end{document}